\documentclass[a4,traditabstract]{aa} 
\usepackage{epsfig}
\usepackage{graphicx}
\usepackage{epstopdf}
\usepackage{color}  
\usepackage{array}   
\usepackage{units}  
\usepackage[breaklinks, colorlinks, citecolor=blue]{hyperref}
\usepackage{natbib}  
\usepackage{multirow}   
\usepackage{amsfonts}
\usepackage{amsmath,amssymb}             
\usepackage[english]{babel}
\usepackage[latin1]{inputenc}
\usepackage[T1]{fontenc}
\usepackage{longtable,lscape}

\def\reff@jnl#1{{\rm#1\/}}
\def\apj{\reff@jnl{ApJ}}       
\def\apjs{\reff@jnl{ApJS}}     
\def\apjl{ApJ}%
\def\ao{Appl.~Opt.}%
\def\aaps{\reff@jnl{A\&AS}}    
\def\aaps{A\&AS}%
\def\mnras{\reff@jnl{MNRAS}}   
\def\prd{\reff@jnl{Phys.\ Rev.\ D}}    
\newcommand{\unb}{\boldsymbol{1}}  
\newcommand{\zerob}{\boldsymbol{0}}  
\newcommand{\Ab}{{\cal{A}}}  
\newcommand{\Sb}{{\boldsymbol{S}}}  
\newcommand{\Tb}{{\boldsymbol{T}}}  
\newcommand{\Cb}{{\cal{C}}}  
\newcommand{\Nb}{\bold{N}}  
\newcommand{\fb}{\boldsymbol{f}}  
  
\newcommand{\wb}{\boldsymbol{w}}  
\newcommand{\eb}{\boldsymbol{e}}  
\newcommand{\cb}{\boldsymbol{c}}  
\newcommand{\nablab}{\boldsymbol{\nabla}}  
\newcommand{\lambdab}{\boldsymbol{\lambda}}

\newcommand{\Db}{{\cal{D}}}

\newcommand{\Ub}{{\cal{U}}}  
  
\newcommand{\gb}{\boldsymbol{g}}  
  
\newcommand{\be}{\begin{equation}}  
\newcommand{\ee}{\end{equation}}  
\newcommand{\bea}{\begin{eqnarray*}}  
\newcommand{\eea}{\end{eqnarray*}}

\begin{document}
\title{MILCA, a Modified Internal Linear Combination Algorithm to extract astrophysical emissions from multi-frequency sky maps.}
\author{G. Hurier\inst{1,2} 
\and J. F. Mac\'{\i}as-P\'erez\inst{1}
\and S. Hildebrandt \inst{1,3}}   
\institute{Laboratoire de Physique Subatomique et de Cosmologie, Universit\'{e} Joseph Fourier Grenoble I, CNRS/IN2P3, Institut National Polytechnique de Grenoble, 53 rue des Martyrs, 38026 Grenoble cedex, France
\and
Institut d'Astrophysique Spatiale, CNRS (UMR8617) Universit\'{e} Paris-Sud 11, B\^{a}timent 121, Orsay, France
\and Jet Propulsion Laboratory, California Institute of Technology, 4800 Oak Grove Drive, Pasadena, California, U.S.A.
}  
\date{Received <date> / Accepted <date>}  

  \abstract
{The analysis of current Cosmic Microwave Background (CMB) experiments is based on the interpretation of 
multi-frequency sky maps in terms of different astrophysical components and it requires specifically tailored component separation algorithms.
In this context, Internal Linear Combination (ILC) methods have been extensively used to extract the CMB emission from the
WMAP multi-frequency data.  We present here a Modified Internal Linear Component Algorithm (MILCA)
that generalizes the ILC approach to the case of multiple astrophysical components for which the
electromagnetic spectrum is known.
In addition MILCA corrects for the intrinsic noise bias in the standard ILC
approach and extends it to an hybrid space-frequency representation of the data. 
It also allows us to use external templates to minimize the contribution of extra components
but still using only a linear combination of the input data.
We apply MILCA to simulations of the \textit{Planck} satellite data at the HFI frequency bands.
We explore the possibility of reconstructing the Galactic molecular CO emission on the \textit{Planck} maps
as well as the thermal Sunyaev-Zeldovich effect.
We conclude that MILCA is able to accurately estimate those emissions and it has been successfully used for this purpose within the Planck collaboration.
               }
                        
\authorrunning{Hurier, Mac\'{\i}as-P\'erez \& Hildebrandt}
\titlerunning{MILCA}

   \maketitle
%

\section{Introduction}

Multi-frequency Comic Microwave Background (CMB) experiments such as WMAP \cite{bennett} and Planck \cite{planck} spacecraft missions, observes a mixture of astrophysical emissions. Most of them follow different emission laws (e.g. CMB anisotropies, thermal Sunyaev-Zel'dovich (tSZ) effect \cite{SZ}, thermal dust, Cosmic Infrared Background (CIB), synchrotron, molecular cloud emission, etc.).
It is thus possible to separate the mixed astrophysical sources taking advantage of the specific emission laws of each astrophysical component.
The separation of astrophysical sources from a set of multi-channel observations of the millimeter and sub-millimeter sky is an important step in the scientific exploitation of such data. In this context, a large number of methods have been developed.\\
These component separation techniques can be separated in various categories depending on the degree and nature of the prior information used.    
For example, Maximum Likelihood techniques \citep[e.g., Commander,][]{eriksen2} and Maximum of Entropy methods (e.g. \citep{fastmem,fastmem2})   
assume that the electromagnetic spectrum of the different components is known or that it can be easily parametrized.  By contrast,
blind (or semi-blind) component separation methods such as, Spectral Matching ICA~\citep{smica,smica2}, 
PolEMICA~\citep{compsep_polemica}, FastICA~\citep{fastICA}, CCA~\citep{bonaldi} or GMCA~\citep{bobin} exploit the spectral
and spatial diversity of the components and need no \emph{a priori} on those. In a different way, ILC (Internal Linear Combination) techniques
 \citep{bennett,eriksen,remazeilles} aim at preserving an astrophysical component for which the electromagnetic
 spectrum is known, by minimizing the variance in the reconstructed signal.  \\
Standard ILC methods are very popular because of  their simplicity and robustness. 
However, as discussed by \cite{vio}, they are biased by the noise contribution to the variance of the final map.
This bias reduces the efficiency of the algorithm in poor signal to noise conditions. 
Furthermore, they do not fully exploit 
the spatial and spectral diversity to minimize the contribution from other sky components 
for which the electromagnetic spectrum is poorly, or not at all, known.\\
The ILC methods also assume that the component of interest is statistically independent from the other ones.
If the statistical independence assumption can be made for the CMB, it is not the case for all the astrophysical components of the millimeter and sub-millimeter sky. Indeed, for example galactic foregrounds such as thermal dust, synchrotron and molecular cloud emission are highly correlated over the sky. Such correlations can also be observed for extra-galactic components like tSZ and extra-galactic point sources.\\
We present, in this article the Modified Internal Linear Combination Algorithm (MILCA) that generalizes standard ILC techniques 
to multiple constraints, in order to reduce the level of contamination by other astrophysical emissions. 
We will show that MILCA also corrects for the noise bias of the standard ILC solution and blindly minimizes the contribution from unknown or poorly constrained sky components. \\
The paper is organized as follows. 
In section~\ref{secsim} we present our modeling the millimeter and sub-milimeter sky. Then, in Sect.~\ref{genILC} we address the problem of generalizing the ILC algorithm to multiple constraints. Section~ \ref{ilcbias} discusses possible bias in the ILC algorithm. In Sect.~\ref{secmod} we propose modifications to the standard ILC estimator in order to minimize the 
instrumental noise and contamination from other astrophysical components. Finally we apply MILCA on Planck simulated datasets, focusing on the extraction of the tSZ effect(Sect.~\ref{reconsz}) and molecular cloud emission (Sect.~\ref{reconco}). 

\section{Simulation of the microwave sky seen by Planck}
 \label{secsim}
 
 \subsection{The \textit{Planck} mission and the sub-millimeter sky}
 The \textit{Planck} mission \citep{planck2011-1.1,planck2013-p01} is the third space mission, 
after the COBE \citep{1992ApJ...396L...1S,bennett2} and  the WMAP \citep{hinshaw1,hinshaw2,bennett} missions,
devoted to the study of the microwave and sub-mm emissions, and especially to the Cosmic Microwave Background radiation (CMB).
\textit{Planck}, as other CMB experiments, observes the sky emission in different frequency bands.
In particular, \textit{Planck} has two instruments:  the Low Frequency Instrument (LFI) \citep{bersanelli2010,planck2011-1.4} which has three frequency bands   
centered at 30, 44  and 70 GHz,  and the High Frequency Instrument (HFI) \citep{lamarre2010,planck2011-1.5} which has six frequency bands
centered at 100, 143, 217, 353, 545 and 857 GHz.
The large number of detectors and their high sensitivity, make the \textit{Planck} mission the first CMB experiment expected to be limited
by our knowledge of the foreground emissions rather than by instrumental noise.  \\  
At the \textit{Planck} frequencies the main astrophysical emissions are: the diffuse Galactic free-free, synchrotron, and thermal dust \citep{planck} emissions; the
anomalous microwave emission \citep[AME,][]{planck2011-7.2} ; the molecular Galactic emissions \citep[mainly  $^{12}$CO  in the 100, 217 and 353~GHz bands,][]{planck2013-p03a},  
the emission from Galactic and extra galactic point sources \citep[radio and infrared sources,][]{planck2012-VII,planck2011-1.10,planck11b}; and the thermal
Sunyaev-Zeldovich effect \citep{SZ} in clusters of galaxies~\citep{planck11d,planck2013-p05a}. \\

\subsection{Simulations of the sub-millimeter sky}
We have simulated the microwave and sub-millimeter sky observed by \textit{Planck} using a set of template maps to reproduce astrophysical components.
Within those components we include CMB, diffuse Galactic emissions (synchrotron, free-free and thermal dust), Galactic CO emission,  Galactic and extra-galactic point sources,
CIB and tSZ emissions.  We did not include in the simulation the Anomalous Microwave Emission (AME) for which a template is not available and
that we expect to be subdominant for the following analyses.\\
The CMB was simulated using a random gaussian field as defined by the Planck best-fit CMB angular power spectrum~\citep{cmbspec,planckcosmo}. 
The thermal dust emission was modeled using the reprocessed IRAS maps at 100 $\mu$m \citep[IRIS][]{miv2005} extrapolated to the Planck frequencies assuming a grey body emission law \citep[e.g.][]{des90}. with constant temperature, $T_{\mathrm{dust}}=18$~K) and spectral index $\beta_{\mathrm{dust}} = 1.8$.\\
The synchrotron emission was simulated using the 408 MHz full-sky map from \citet{has81} corrected for free-free emission assuming an emission law with a spatially constant spectral index of $-3.0$ in antenna-temperature units \citep{dav96}. 
The Galactic free-free emission was modeled using the $H_{\alpha}$ full-sky map \citep{fin03} and an electron temperature of $7000$~K as proposed by \citet{dic03} and a spatially constant spectral index of $-2.1$ in antenna-temperature units.
The CO emission was modeled using the J=1-0 CO map from \citet{dam01}, the transmission in Planck's sky maps have been computed assuming a fluctuation of 10\% in the bandpass amplitude \citep{planckco}. Finally, we add gaussian white noise in the map with the expected levels for the \textit{Planck} mission \citep{planck2013-p01}.
\\ 
We also add radio point sources using the sources in the NVSS catalog and extrapolating their flux to the Planck frequencies assuming a random spectral index for each source 
derived from a Gaussian distribution with a mean of $-2.5$ and a standard deviation of  $0.2$  in antenna temperature units \citep{planck11a}. \\
Extra-galactic infra-red (IR) astrophysical emissions can be decomposed into two categories, a poissonian contribution, composed by point sources, and a diffuse contribution, the clustered CIB. To model IR point sources we also used a randomly distributed sources with random spectral index for each source (with a mean of $1.0$ and a standard deviation of  $0.3$  in antenna temperature units \citep{planck11a}). For the CIB emission, we assumed a random gaussian distribution following the CIB power spectra as measured by Planck~\citep{planck11b} and constant correlation factors across frequencies.\\
The tSZ emission is produced by the hot and dense gas of electrons, such gas can be found in galaxy clusters or in the WHIM. In this study we only account for the emission from X-ray known galaxy clusters. To do so, the tSZ  was simulated using a universal electron pressure profile from \citet{arn10}  and 
physical parameters for clusters from the MCXC catalog \citep{pif10}.\\

\section{Generalising ILC method}   
\label{genILC}
The total emission observed at a given frequency channel map, $T_{i}$, can be expressed as the superposition
of astrophysical components and instrumental noise
\begin{equation}
T_{i}(\mathrm{p}) = \sum_{j=1}^{N_{\mathrm{s}}}  {\cal A}_{i j} (\mathrm{p}) \ S_{j}  (\mathrm{p}) + n_{j}(\mathrm{p}),
\label{mdmsp}
\end{equation}
where $\mathrm{p}$ denotes the pixel number. $S_{j}$ are the maps of the $N_{\mathrm{s}}$ astrophysical components (CMB and foregrounds)
in the data and $n_{j}$ is the instrumental noise. ${\cal A}_{i j}$ is the mixing matrix given by
\be
{\cal A}_{ij} = \int \,F_j\,(\nu)\, H_i\,(\nu) \, {\mathrm{d}}\nu,
\label{freqdep}   
\ee
with $F_j\,(\nu)$ the spectral dependence of the $j$-th component and $H_i\,(\nu)$ the spectral response of the $i$-th channel.  \\

Assuming that the mixing matrix is constant across the sky, equation~\ref{mdmsp} can be written in a more compact manner as
\be  
\Tb (\mathrm{p}) = \Ab  \Sb (\mathrm{p}) + \Nb(\mathrm{p}),  
\label{modelling}   
\ee   
where $\Tb~$ and $\Nb$ are vectors containing the total and noise maps for the $N_{\mathrm{obs}}$ observation channels,
 $\Sb $ is a matrix corresponding to the maps of the $N_{\mathrm{s}}$ astrophysical components. $\Ab$ is  the mixing matrix with
dimensions $N_{\mathrm{obs}}~\times~N_{\mathrm{s}}$.

\subsection{Internal Linear Combination algorithm}
\label{ilcalgo}
The Internal Linear Combination algorithm \citep{bennett} assumes that the astrophysical component to be reconstructed, $S_{\mathrm{c}}$, can be obtained as a linear combination
of the input observational maps
\be  
\hat{S}_{\mathrm{c}} (\mathrm{p})  = {\wb}^{\Tb} \Tb  = \sum_{i} w_{i}T_{i}  (\mathrm{p}).
\label{objectif}   
\ee   
The weights, $\wb^{\Tb} = (w_{1}, \dots, w_{N_{\mathrm{obs}}})$, of the linear combination are computed by minimizing the variance of the output map, 
\begin{align}
\nonumber
Var(\hat{S}_{\mathrm{c}}) & =  \frac{1}{N_{\mathrm{pix}}} {\sum_{\mathrm{p}=1}^{N_{\mathrm{pix}}} \hat{S}_{\mathrm{c}}^{2} (\mathrm{p})  -  {\left(\sum_{\mathrm{p}=1}^{N_{\mathrm{pix}}} \hat{S}_{\mathrm{c}} / N_{\mathrm{pix}} \right)}^{2}} \\
& =   {\wb}^{\Tb} \left <  \Tb  \Tb^{\Tb} \right > \wb =  {\wb}^{\Tb}  \Cb_{\Tb}  \wb,
\end{align}
under constraints on the emission spectrum of $S_{\mathrm{c}}$. $\Cb_{\Tb}$ is the covariance
matrix of the observed maps, $\Tb$, averaged across pixels. \\

For the CMB and assuming that the maps are given in thermodynamic temperature units, the constraint simply read \citep{bennett,eriksen}
\be
g = \sum_i {\wb}_i = 1.
\ee
More generally, for an astrophysical component with emission spectrum, $\fb_{\cb}$, the constraints can be written as
\be   
g = {\wb}^{\Tb} \fb_{\cb} = 1.   
\label{flatcons}   
\ee
Notice that $\fb_{\cb}$ is related to the mixing matrix, $\Ab$, by $\fb_{\cb} = \Ab \eb_{\cb}$, where $\eb_{\cb}$ is an $N_{\mathrm{s}}$ dimension vector 
for which all elements are set to zero except the $c$-th element that is set to 1.

\subsubsection*{Linear system solution}
Weights can be computed solving a linear system using Lagrange multipliers
\begin{align} 
\nablab V(S_{\mathrm{c}}) - \lambda \nablab g &= 0  \nonumber \\      
\gb = \fb_{\cb}^{\Tb} \wb &= 1  ,
\label{eqflat}   
\end{align}
where $\lambda$ is a Lagrange multiplier.
The final linear system can be expressed in the form
\begin{align}
\left( \begin{array}{cc}   
2.\Cb_{\Tb} & -\fb_{\cb} \\   
{\fb_{\cb}}^{\Tb} & \zerob    
\end{array} \right)   
\left( \begin{array}{c}   
\wb  \\   
\lambda   
\end{array} \right)   
=\left( \begin{array}{c}   
\zerob   \\   
1   
\end{array} \right).   
\label{newsys}   
\end{align}

\noindent Solving the linear system we obtain
\be   
\widehat \wb = {\Cb}_{\Tb}^{-1} {{\fb_{\cb}}} \left( {{\fb_{\cb}}}^{T}  {\Cb}_{\Tb}^{-1} {\fb_{\cb}} \right)^{-1} 
\label{westflat}   
\ee 
and then
\be   
\widehat S_{\mathrm{c}} (\mathrm{p})= \left({ {\fb_{\cb}}^{T}  {\Cb}_{\Tb}^{-1} \fb_{\cb} } \right)^{-1} {\fb}_{\cb}^{T}  {\Cb}_{\Tb}^{-1} \Tb (\mathrm{p}).
\label{scestflat}   
\ee  
In the case of the CMB emission Eq.~\eqref{scestflat} reduces to 
\be   
\widehat S_{\mathrm{c}} (\mathrm{p}) = \frac{  \unb^{T} {\Cb}^{-1}_{\Tb} \Tb(\mathrm{p})}{{\unb}^{T} {\Cb}^{-1}_{\Tb} \unb},   
\label{classic}   
\ee   
where $\unb = (1, \dots, 1)$, as in \citet{bennett} and \citet{eriksen}.

\subsection{Multiple spectral constraints}
\label{multiplespectralconstraints}

In some particular cases, the emission spectra of various astrophysical components may be known with sufficient accuracy. 
Thus, it is possible to use multiple constraints (i.e., $g_{1}, g_{2}, \dots, g_{N_{\mathrm{nc}}}$) for the minimization process to
extract some of the astrophysical components and to reject some of the others.
In practice, we can write those constraints as
\begin{align}
g_{1}  =   {\wb}^{\Tb} \fb_{1} = 1 \nonumber \\  
g_{2} = {\wb}^{\Tb} \fb_{2} = 0  \nonumber \\ 
\vdots \nonumber \\
g_{N_{\mathrm{rc}}+1} = {\wb}^{\Tb} \fb_{N_{\mathrm{rc}}+1} = 0,
\label{multiplecons}   
\end{align}
where $N_{\mathrm{rc}}$ are the number of rejected components. \\

\noindent 

\citet{remazeilles} has shown that 2 components can be constrained and simultaneously recovered. We propose here a generalization of this method to 
an arbitrarily large number of constraints.
Defining $\cal F$ as a matrix with $N_{\mathrm{obs}} ~\times~ (1 + N_{\mathrm{rc}})$ elements such that
{\tiny
 $$
 \cal F = 	
  \begin{pmatrix}
  \fb_{1}[1]                                & \fb_{2}[1] 				& \cdots &   \fb_{1+N_{\mathrm{rc}}}[1] \\
  \vdots                                      & \vdots  							&\ddots  &  \vdots  		\\
  \fb_{1}[N_{\mathrm{obs}}]   &  \fb_{2}[N_{\mathrm{obs}}]	& \cdots  &  \fb_{1+N_{\mathrm{rc}}}[N_{\mathrm{obs}}] \\
 \end{pmatrix},
 $$
 }
 equation \ref{newsys} can be generalized to 
 \begin{align}  
\left( \begin{array}{cc}   
2.\Cb_{\Tb} & -{\cal F} \\   
{\cal F}^{\Tb} & \zerob    
\end{array} \right)   
\left( \begin{array}{c}   
\wb  \\   
\lambdab  
\end{array} \right)   
=\left( \begin{array}{c}   
\zerob   \\   
\eb 
\end{array} \right).   
\label{systmultconst}   
\end{align}
where $\lambdab=\left(\lambda_{(1)}, \lambda_{(2)}, \cdots, \lambda_{(1+N_{\mathrm{rc}})}\right)^{T}$ are the Lagrange multipliers and $\eb=\left(1,0,\dots,0\right)^{T}$.
The solution of the system can then be written as
\be   
\widehat S_{\mathrm{1}} (\mathrm{p}) = \eb^{\Tb} {\cal W}^{\Tb} \Tb(\mathrm{p}) =  \lbrace \eb^{T}  \left({ {\cal F}^{T}  {\Cb}_{\Tb}^{-1} \cal F } \right)^{-1} {\cal F}^{T}  {\Cb}_{\Tb}^{-1} \rbrace \Tb (\mathrm{p}),
\label{weightsolmultconst}   
\ee  
with ${\cal W}$ the $(N_{\mathrm{rc}}+1)~\times~N_{\mathrm{obs}} $ dimension matrix containing the weights extracting each constrained components. The first line of this matrix contains the weights for the extraction of the components of interest, $\wb = {\cal W} \eb$.\\
Notice that in the relation ${\cal F} = \Ab \eb_{\cb}$, $\eb_{\cb}$ is now a $N_{\mathrm{s}}~\times~ (1 + N_{\mathrm{rc}})$ matrix.

\section{Bias in the standard ILC estimator}
\label{ilcbias}
The ILC estimator as presented above is biased. This bias can have several origins:
\begin{itemize}   
\item correlation between the wanted components and the other astrophysical components
\item differences between the \emph{true} emission laws and the constraints used to minimize the variance. These differences can be produced by calibration uncertainties \cite{dick} or by discrepancies between the used emission laws and the effective ones.
\item noise-induced correlations
\end{itemize}   

\subsection{Intrinsic bias }
\label{secbias}
We first characterize the bias in the ILC estimator proposed in Eq.~\eqref{westflat}, neglecting the contribution of the instrumental noise. 
In this situation, we can write the channel covariance matrix as:
\be   
{\Cb}_{\Tb} = \Ab \Cb_{\Sb} \Ab^{\Tb},   
\label{ctnoiseless}   
\ee   
with $\Cb_{\Sb}$ the $N_{\mathrm{s}}~\times~N_{\mathrm{s}}$ dimension covariance matrix of the astrophysical components. 
The ${\Cb}_{\Tb}$ matrix has a dimension $ N_{\mathrm{obs}}~\times~N_{\mathrm{obs}} $ but has a rank: $ {\mathrm{min}}(N_{\mathrm{s}},N_{\mathrm{obs}}) $, by construction; as for $\Cb_{\Sb}$, it has a rank $N_{\mathrm{s}}$. 
From here, three cases are possible for the ${\Cb}_{\Tb}$ matrix inversion:

\begin{itemize}   
\item (1) $ N_{\mathrm{s}} > N_{\mathrm{obs}} $:  There are more physical components than channels.  It is not possible to describe ${\Cb}_{\Tb}$ with an $n_{\mathrm{s}}$ dimension sub-space. The $n_{\mathrm{t}}$ dimension subspace which describe the ${\Cb}_{\Tb}$ matrix has thus no physical meaning and consequently it is not possible to extract a single component by a linear combination. The ILC approach will thus produce highly biased estimation.  A solution to this problem requires more information, such as other channels of observations.
\item (2) $N_{\mathrm{s}} = N_{\mathrm{obs}}$: There are the same number of components than channels. In this case, $\Ab$ is a square matrix. There is no ambiguity in inverting ${\Cb}_{\Tb}$. There exist an unbiased linear combination that allows us to extract $S_{\mathrm{1}}$. 
\item (3) $N_{\mathrm{s}} < N_{\mathrm{obs}}$: There are more channels than astrophysical components. ${\Cb}_{\Tb}$ has a rank $N_{\mathrm{s}}$ lower than his dimension $N_{\mathrm{obs}}$. Consequently, this matrix is singular and cannot be directly inverted. However, we can use the \emph{pseudo inverse} as defined by the Singular Value Decomposition (SVD) of the matrix. We can thus write ${\Cb}_{\Tb}^{-1} = \Ub {\Db}^{-1} {\Ub}^{\Tb}$, with $\Ub$ an orthogonal matrix and $\Db$ a diagonal matrix containing the singular values of ${\Cb}_{\Tb}$. The ${\Db}^{-1}$ matrix is obtained by taking the inverse of all singular values different from zero, and setting to zero the inverse of all singular values equal to zero. The $\Db$ matrix is uniquely defined, however the $\Ub$ matrix is defined with $N_{\mathrm{obs}} - N_{\mathrm{s}}$ degrees of freedom. There exist multiple linear combinations that can extract the $S_{\mathrm{1}}$ component.
\end{itemize}   
When the $\Ab$ matrix is rectangular ($N_{\mathrm{s}} < N_{\mathrm{obs}}$), the notation $\Ab^{-1}$ refers to the left inverse of the matrix (also defined with $N_{\mathrm{obs}} - N_{\mathrm{s}}$ degrees of freedom).\\

For each pixel we can also define $\Sb'(p)$ and $\Sb''(p)$ two vectors of dimensions $N_{\mathrm{rc}}+1$ and $N_{\mathrm{s}}-N_{\mathrm{rc}}-1$ respectively, that contain the constrained and un-constrained components such that
\begin{align}  
\Sb(p) &=\left( \begin{array}{c}   
\Sb' (p)\\   
\Sb'' (p)
\end{array} \right).
\end{align}  
For simplicity hereafter we do not explicitly write the pixel index, $p$. \\

In the following, we will focus on the case  $N_{\mathrm{s}} \leq N_{\mathrm{obs}}$.
Using Eq.~\eqref{ctnoiseless}, we can express the estimator of $S_{\mathrm{1}}$ as
\begin{align}   
\widehat  S_{\mathrm{1}} & =  \eb^{\Tb} ({\eb}^{\Tb}_{c} {\Cb}^{-1}_{\Sb} \eb_c)^{-1} {\eb}^{T}_{c} {\Cb}_{\Sb}^{-1} \Sb.
\label{eq-hatsc} 
\end{align}
To simplify the following discussions, we chose to express ${\Cb}_{\Sb}$ as a function of different sub-spaces
\begin{align}  
{\Cb}_{\Sb} &=\left( \begin{array}{cc}   
{\cal{E}}_{0} & {\cal{G}}_{0} \\   
{\cal{G}}^T_{0} & {\cal{H}}_{0}   
\end{array} \right)
\label{subcs}
\end{align}  
with ${\cal{E}}_{0}$ the $(N_{\mathrm{rc}}+1)~\times~(N_{\mathrm{rc}}+1)$ dimension sub-space associated with the constrained components,  ${\cal{H}}_{0}$ the $(N_{\mathrm{s}}- N_{\mathrm{rc}}-1)~\times~(N_{\mathrm{s}} - N_{\mathrm{rc}}-1)$ dimension sub-space associated with the other components. ${\cal{G}}_{0}$ is a  $(N_{\mathrm{rc}}+1)~\times~(N_{\mathrm{s}} - N_{\mathrm{rc}}-1)$ sub-space contains the correlation between the contained and un-constrained components.\\

We can rewrite the estimator of $S_{\mathrm{1}}$, using a bloc inversion procedure
\begin{align}   
\widehat  S_{\mathrm{1}} & =  \eb^{\Tb}\left(\Sb' - {\cal{G}}_0 {\cal{H}}^{-1}_0 \Sb''\right) \nonumber \\
                         & =  S_{\mathrm{1}} - \eb^{\Tb}{\cal{G}}_0 {\cal{H}}^{-1}_0 \Sb''.
\label{biasnonoise}
\end{align}
We note that the $S_{\mathrm{1}}$ component is effectively recovered. However, we also notice the presence of a term of bias $- \eb^{\Tb} {\cal{G}}_0 {\cal{H}}^{-1}_0 \Sb''$ produced by the correlation, ${\cal{G}}_0$, between the reconstructed component and the un-constrained components. 
This term of bias is composed by a linear combination of the un-constraints components and it is also function of the inverse of the covariance matrix 
of the un-constrained component. \\
This bias can be intuitively understood as the astrophysical emissions correlated with the wanted one, $S_{\mathrm{1}}$, can be used to minimize the variance by removing a fraction of the wanted emission. In order to reduce the bias induced by such correlations, it is possible to use prior information on the spatial distribution of the contaminating astrophysical components as discussed in~Sect.~\ref{subsect:p2c1:tilt}.

\subsection{Noise-induced bias}  
\label{sec-noise}  
We now focus on the impact of the noise on the estimation of the wanted component, $S_{\mathrm{1}}$.
Adding noise, the ${\Cb}_{\Tb}$ matrix can be simply written as
\be   
{\Cb}_{\Tb} = \Ab \Cb_{\Sb} \Ab^{T} + \Cb_{\Nb},
\label{ctnoised}   
\ee  
where ${\Cb}_{\Nb} = \langle \Nb {\Nb}^{T} \rangle$ is the $N_{\mathrm{obs}}~\times~N_{\mathrm{obs}}$ dimension covariance matrix of the instrumental noise averaged across pixels. This matrix is diagonal if the noise between different observation channels is uncorrelated. 
This matrix has a rank $N_{\mathrm{obs}}$, consequently ${\Cb}_{\Tb}$ has also rank $N_{\mathrm{obs}}$ and can be inverted in any case.
The ${\Cb}_{\Tb}$ matrix can be projected in the space of the astrophysical components as 
\be   
\Ab^{-1} {\Cb}_{\Tb} \Ab^{-\Tb} =  \Cb_{\Sb}  +  \Ab^{-1} \Cb_{\Nb}  \Ab^{-\Tb}.   
\label{csnoise}   
\ee   
Assuming that  ${\Cb}_{\Nb}$ is a diagonal matrix (this is the case for the applications presented in this paper), 
$\Ab^{-1} \Cb_{\Nb}  \Ab^{-\Tb}$ is not.
The noise covariance matrix adds non-diagonal terms in $\Ab^{-1} {\Cb}_{\Tb} \Ab^{-\Tb}$. As discussed in Sect.~\ref{secbias}, this term will add a bias in the estimator of $\widehat S_{\mathrm{1}}$ such that
\begin{align}   
\widehat S_{\mathrm{1}}  =&  \eb^{\Tb} ({\cal F}^{T} {\Cb}^{-1}_{\Tb} {\cal F})^{-1}{\cal F}^{T} {\Cb}^{-1}_{\Tb} {\Tb}. 
\label{eq-noisebias0}  
\end{align}   
Writing  ${\Tb} = \Ab \Sb + \Nb$ (signal plus noise terms) equation~\ref{eq-noisebias0} reads
{\tiny
\begin{align}   
\widehat S_{\mathrm{1}} =& \eb^{\Tb}({\eb}_{c}^{T}  ({ \Cb}_{\Sb} + \Ab^{-1} { \Cb}_{\Nb} {\Ab}^{-\Tb} )^{-1} \eb_{c})^{-1}{\eb}_{c}^{T}({ \Cb}_{\Sb} + \Ab^{-1} { \Cb}_{\Nb} {\Ab}^{-\Tb})^{-1} \Sb \nonumber \\
 &   + \eb^{\Tb} ({\eb}_{c}^{T}  ({ \Cb}_{\Sb} + \Ab^{-1} { \Cb}_{\Nb} {\Ab}^{-\Tb} )^{-1} \eb_{c})^{-1}{\eb}_{c}^{T} \Ab^{\Tb} (\Ab { \Cb}_{\Sb} {\Ab}^{\Tb} + { \Cb}_{\Nb})^{-1} \Nb. \qquad 
\label{eq-noisebias}  
\end{align}   
}
We observe two terms, one depending on the astrophysical signal and the second depending on the noise contribution.
To express clearly the extra-bias produced by noise, we rewrite the matrix $ \Ab^{-1} \Cb_{\Nb}  \Ab^{-\Tb}$ in the same form we used for $\Cb_{\Sb}$ in Eq~\ref{subcs} 
\begin{align}  
{\Ab}^{-1} \Cb_{\Nb} \Ab^{-\Tb} &=\left( \begin{array}{cc}   
{\cal{E}}_{N} & {\cal{G}}_{N} \\   
{\cal{G}}^{T}_{N} & {\cal{H}}_{N}   
\end{array} \right)
\end{align}  
and
\begin{align}  
\Ab^{-1}\Nb &=\left( \begin{array}{c}   
\tilde \Nb' \\   
\tilde \Nb'' 
\end{array} \right).
\end{align}  
We can write the residual term in our estimator in the form:
{\tiny
\begin{align}   
R_{1} &= \widehat  S_{\mathrm{1}} - S_{\mathrm{1}} \nonumber \\
&=  \eb^{\Tb} \tilde{\Nb'} - \eb^{\Tb} ({\cal{G}}_0 + {\cal{G}}_N) ({\cal{H}}_0 + {\cal{H}}_N)^{-1} (\Sb'' + \tilde \Nb'') \nonumber \\
&= \eb^{\Tb} \tilde{\Nb'} - \eb^{\Tb} ({\cal{G}}_0 + {\cal{G}}_N) {\cal{H}}^{-1}_0 \left[\sum_{k=0}^{\infty} (-1)^{k} ({\cal{H}}_0^{-1}{\cal{H}}_N)^{k}\right] (\Sb'' + \tilde \Nb'').
\label{biaswithnoise}
\end{align}
}
The term ${\cal{G}}_N$ produces extra-bias. The extra bias introduced by the noise can be reduced if we  have an estimate of the noise covariance matrix, $\Cb_{\Nb}$.
We will discuss the correction of this bias and its caveats in Sect.~\ref{secmod}.
    
\section{Modified Internal Linear Combination Algorithm (MILCA)}  
\label{secmod}
To reduce the bias presented in the previous section, we propose several modifications to the standard ILC estimator:

\begin{itemize}
\item (1) localization in pixel and spherical harmonic spaces to account for spatial spectral law variations 
\item (2) modify the definition of the variance we minimize (by an action on the covariance matrix, which is equivalent to a modification of the Lagrangian of the problem),
to account for noise-induced and astrophysical correlations
\item (3) add extra constraints as discussed in Sect.~\ref{multiplespectralconstraints}.
\end{itemize}

\subsection{Localization in the pixel and spherical harmonic domains}
\label{subsect:p2c1:filtre}

Astrophysical components properties vary both spatially (pixel domain) and in frequency (spherical harmonic domain). For example, the CMB is homogenous and isotropic over the sky and is dominant at a typical angular scale of about one degree. Galactic foregrounds are localized in the galactic plane and at large angular scales with spectral laws
that vary smoothly spatially. Extra-galactic foregrounds are localized at small angular scales, with emission laws changing significantly from an object to another. Consequently using a reconstruction localized both in space and frequency allows to improve the ILC performances, by adapting the weights $\wb$ to the local background. This point has intensively been discussed in the literature (e.g. \cite{nilc,gmca}) .\\

We choose here to filter the observations with $N_{k}$ filters in spherical harmonics space. These filters are built from the difference between two Gaussian filters
of the form $B^{\alpha}_{l} = {\mathrm{exp}}(-l(l+1)\sigma_{\alpha}^2/2)$ such that
\be
F_{l}^{\alpha} = B_{l}^{\alpha} - B_{l}^{\alpha+1}
\ee
where $\sigma^2_{\alpha}$ increases with $\alpha$. We also impose $B_{l}^{0} = 1$ et $B_{l}^{k} = 0$, in order that the condition
\be
\sum^{k-1}_{\alpha} F_{l}^{\alpha} = 1,
\ee
is satisfied.\\
After filtering the observation channel maps, the weights $\wb$ are computed locally in predefined pixel regions.
To ease the procedure these regions have been built using the properties of the \textit{HEALPix} pixelisation NESTED scheme \citep{healpix}.
For each filter, these regions are defined by \textit{HEALPix} pixels at a given resolution $N^{\alpha}_{side}$ such that
the size of the pixel is 20 times greater than $\sigma_{\alpha+1}$ and 
$N^{\alpha}_{side} \leq \frac{N_{side}}{16}$, where $N_{side}$ is the native resolution of observation channel maps.
In order to ensure the continuity of the weights, $\wb$, at the interface between two contiguous pixel regions at the resolution $N^{\alpha}_{side}$, the map of weights at the native resolution $N_{side}$ are convolved with a Gaussian beam with a FWHM equal to the size of the HEALPIX pixel at the resolution $N^{\alpha}_{side}$.\\

\subsection{External template regularization of the ILC solution}
\label{subsect:p2c1:tilt}

In the following we consider the possibility of using external templates to minimize the contribution of unwanted components, particularly those
for which the electromagnetic spectrum is poorly known.
For example point-like sources and compact objects have varying electromagnetic spectra and therefore are difficult to handle with the standard ILC algorithm.\\
One of the main advantages of ILC methods is to only use information from a single experiment in the linear combination. 
We present here an approach allowing us to use external priors for the computation of the weights $\wb$ but not using the external templates themselves in the linear combination.
To do so, we modify the data covariance matrix, ${\Cb}_{\Tb}$ including an extra term.
Given unwanted astrophysical components we first compute their expected contribution to the data covariance matrix and then
we \emph{exacerbate} it to force the ILC algorithm to minimize them in the final estimate of $\widehat S_{\mathrm{1}}$.
In practice we write
\be
{\Cb}_{\Tb}' = \Cb_{\Tb} + (\gamma^2-1){\Cb}_{C}
\label{modifcovmatrix}
\ee
$\Cb_{C}$ a $N_{\mathrm{obs}}~\times~N_{\mathrm{obs}}$ dimension matrix containing the contribution of the unwanted astrophysical components
to the data covariance matrix as estimated from external templates of these components. 
$\gamma$ is a multiplicative factor that takes only values larger than 1 and it is adapted depending on the accuracy of the estimated $\Cb_{C}$. 
Writing $\Cb_{C}$ as
\begin{align}  
\Cb_{C} &=\Ab \left( \begin{array}{cc}   
{{0}} & {{0}} \\   
{{0}} & {\cal{H}}_{C}   
\end{array} \right) \Ab^{\Tb},
\end{align}  
we can derive the following expression for the residuals
{\tiny
 \begin{align}   
R_{1} &= \eb^{\Tb} \left[ \tilde{\Nb'} - ({\cal{G}}_0 + {\cal{G}}_N) \left({\cal{H}}_0 + (\gamma^2-1){\cal{H}}_C + {\cal{H}}_N\right)^{-1} (\Sb'' + \tilde \Nb'') \right] \nonumber \\
&= \eb^{\Tb} \left[ \tilde{\Nb'} - ({\cal{G}}_0 + {\cal{G}}_N) \left({\cal{H}}_0' + {\cal{H}}_N\right)^{-1} (\Sb'' + \tilde \Nb'') \right].
\label{biaswithnoise2}
\end{align}
}
The ${\cal{H}}_C$ matrix is in general singular as all the astrophysical emissions are not constrained. As a consequence this matrix only modify a sub-space of ${\cal{H}}_0$. \\

In Sect.~\ref{seccib} we present a practical application of this technique in order to reduce the CIB contamination in the reconstructed 
tSZ map.  We observe that it leads to a reduction of the bias in the residuals proportional to the factor $\simeq \gamma^2$,
but also to an increase of the noise.

\subsection{Reducing the noise-induced bias}
\label{secrmbias}
As discussed in Sect.~\ref{sec-noise} instrumental noise produces a bias in the estimate of $S_{\mathrm{1}}$. 
To remove this bias, \cite{vio} proposed an unbiased estimator that  can be obtained by modifying the data covariance matrix as follows 
\begin{align}   
\widetilde{{\Cb}}_{\Tb} =  {\Cb}_{\Tb}' - \widehat{\Cb}_{\Nb}. 
\label{unbiasedest}   
\end{align}   
If the direct subtraction of the noise in $\Tb$ is not possible, the estimation of the noise covariance matrix $\Cb_{\Nb}$ is possible.
The estimate of the data covariance matrix in  Eq.~\eqref{unbiasedest} allows us to obtain an estimate of $S_{\mathrm{1}}$ 
equivalent to the one for the noiseless case, as presented in Sect.~\ref{secbias}. 
Consequently, $\widehat S_{\mathrm{1}}$ is only biased by the correlation between the wanted, $S_{\mathrm{1}}$, and the other components.
Indeed, we obtain the following expression for the residual
\be
R_{1} = \widehat S_{\mathrm{1}} - S_{\mathrm{1}} =  \eb^{\Tb} \left[ \tilde \Nb' - {\cal{G}}_{0} {{\cal{H}}'}_{0}^{-1} (\Sb''+\tilde \Nb'')\right].
\label{restilt}
\ee
The excess of bias introduced by the noise covariance matrix has been suppressed. \\

However, in this situation we do not longer minimize the contribution of the noise to the data covariance matrix ${\Cb}_{\Tb}$. 
Furthermore, the ${\widetilde \Cb}_T $ matrix is singular for the case $N_{\mathrm{s}} < N_{\mathrm{obs}}$. 
This can lead to a significant increase of the instrumental noise in $S_{\mathrm{1}}$. 
To balance this effect and minimize the noise contribution we add a regularization term to the ILC algorithm. 
We use the remaining $N_{\mathrm{obs}} - N_{\mathrm{s}}$ degrees of freedom in the definition of the pseudo inverse for the ${\tilde \Cb}_{\Tb}$ matrix. \\

We start by clearly identifying the subspaces of $\widetilde{{\Cb}_{\Tb}}$ that are constrained by the ILC constraints and the variance minimization respectively.
The ILC constraints act on a $N_{\mathrm{rc}}+1$ dimension subspace, then the minimization of the variance act on a $N_{\mathrm{s}} - N_{\mathrm{rc}} - 1$ dimension subspace.
 We still have an $N_{\mathrm{obs}} - N_{\mathrm{s}}$ dimension subspace to minimize the instrumental noise contribution. 

\subsubsection{Constrained-components subspace}
In order to isolate the subspace associated with the ILC constraints, we propose to perform a transformation of the form
\begin{align}
\widetilde{{\Cb}_{\Tb}} &=  {\Cb}_{\Tb}' - {\Cb}_{\Nb} - \alpha {\cal F}( {\cal F}^{\Tb} ({\Cb}_{\Tb}' - {\Cb}_{\Nb})^{-1} {\cal F})^{-1} {\cal F}^{\Tb} \nonumber \\
&=  {\Cb}_{\Tb}' - {\Cb}_{\Nb} - \alpha \Ab \eb_{c} ({\cal{E}}_0 - {\cal{G}}_0 {{\cal{H}}'}^{-1}_0 {\cal{G}}^{\Tb}_0) \eb_{c}^{\Tb} \Ab^{\Tb} \nonumber \\
&=  {\Cb}_{\Tb}' - {\Cb}_{\Nb} - \alpha \Ab \eb_{c} \widehat {\cal{E}}_0 \eb_{c}^{\Tb} \Ab^{\Tb}
\label{unbiasedestp}   
\end{align}
with $\alpha$ the fraction of variance from the constrained components to be subtracted. For $\alpha = 1$ the $\widetilde{{\Cb}_{\Tb}}$ matrix has a rank 
$N_{\mathrm{s}}-N_{\mathrm{rc}}-1$, the ${\Cb}_{\Tb}' - {\Cb}_{\Nb}$ matrix having a rank $N_{\mathrm{s}}$. 
We subtract ${\Cb}_{\Nb}$ because we want to consider the variance of the astrophysical signal without noise. 
$\widehat {\cal{E}}_0 = {\cal{E}}_0 - {\cal{G}}_0 {{\cal{H}}'}^{-1}_0 {\cal{G}}^{\Tb}_0$ is the estimate of the covariance matrix of the constrained components
for the standard ILC algorithm. We observe the contribution of the intrinsic bias term $- {\cal{G}}_0 {{\cal{H}}'}^{-1}_0 {\cal{G}}^{\Tb}_0$. Indeed, 
the variance of the reconstructed component reads
\begin{align}
V_{S_{\mathrm{1}}} &= {\wb}^{\Tb} ({\Cb}_{\Tb}' - {\Cb}_{\Nb}) \wb \nonumber \\
&=  \eb^{\Tb} ({\cal F}^{\Tb} ({\Cb}_{\Tb}' - {\Cb}_{\Nb})^{-1} {\cal F})^{-1} \eb \nonumber \\
&=  \eb^{\Tb} \widehat {\cal{E}}_0 \eb.
\end{align}
Writing the data covariance matrix as
\be
\widetilde{{\Cb}_{\Tb}} = \Ab ({\Cb}_{\Sb}' - \alpha \eb_{c} \widehat {\cal{E}}_0 \eb_{c}^{\Tb}) {\Ab}^{\Tb}.
\label{reduction}   
\ee
and we derive
{\tiny
\be
\widehat S_{\mathrm{1}} = \eb^{\Tb} \left[  {\eb}_{\cb}^{\Tb} ({\Cb}_{\Sb}' - \alpha  \eb_{c} \widehat {\cal{E}}_0 \eb_{c}^{\Tb})^{-1} {\eb}_{\cb}  \right]^{-1} {\eb}_{\cb}^{\Tb}({\Cb}_{\Sb}'- \alpha  \eb_{c}\widehat {\cal{E}}_0 \eb_{c}^{\Tb})^{-1}(\Sb + {\Ab}^{-1}\Nb),
\ee
}
the ${\Cb}_{\Sb}' - \alpha  \eb_{c} \widehat {\cal{E}}_0 \eb_{c}^{\Tb}$ matrix can be rewritten in the following form :
\begin{align}  
{\Cb}_{\Sb}' - \alpha  \eb_{c} \widehat {\cal{E}}_0 \eb_{c}^{\Tb}&=\left( \begin{array}{cc}   
{\cal{E}}_{0} - \alpha \widehat {\cal{E}}_0 & {\cal{G}}_{0} \\   
{\cal{G}}^T_{0} & {\cal{H}}_{0}'   
\end{array} \right).
\end{align}  
Consequently, the residual, $R_1$, in Eq.~\eqref{restilt} do not depend on ${\cal{E}}_0$ and consequently do not depend on the parameter $\alpha$. \\
Indeed, the ${\cal{E}}_{0}$ sub-space is constrained, its modification will not produce any modification on $\widehat S_{\mathrm{1}}$.
We can deduce that the weights $\wb$ are invariant under this transformation. So, we select the value $\alpha = 1$ for which $\tilde{{\Cb}_{\Tb}}$ becomes $N_{\mathrm{rc}}+1$ singular. 

To estimate the value of $N_{\mathrm{s}}-N_{\mathrm{rc}}-1$ components, we compute the number of eigenvalues of the $\tilde{{\Cb}_{\Tb}}$ matrix which are significantly greater than $0$. Notice that in this way we also have an estimate of $N_{\mathrm{obs}}-N_{\mathrm{s}}$ giving the remaining degrees of freedom that can be used to reduce the noise contribution in $\widehat S_{\mathrm{1}}$. 
In the case $ N_{\mathrm{s}} < N_{\mathrm{obs}} $, the pseudo inverse of $\tilde{\Cb}_{\Tb}$ is defined with $N_{\mathrm{obs}}-N_{\mathrm{s}}$ degrees of freedom as explained in the Sect.~\ref{sec-noise}. 

\subsubsection{Minimizing the noise variance}
\label{secsub}
To reduce the noise contribution in the final estimate of $\widehat S_{\mathrm{1}}$ we also
minimize simultaneously the variance of the noise term
\be   
V_N =  \wb^{T} \Cb_{\Nb} \wb \qquad 
\label{noisemin}   
\ee  
This can be performed by modifying the lowest eigenvalues of the $\tilde{\Cb}_{\Tb}$ matrix, assuming that they are not associated to astrophysical emissions. 
We search for the value of the eigenvalues $D_{j}$ of $\tilde{\Cb}_{\Tb} = \Ub \Db \Ub^{\Tb}$ that minimize $V_N$.
The minimization is performed numerically and iteratively. It is important to notice that $V_N$ might have several extrema.
Consequently, we use the first and second derivative of the variance to find the minimum of $V_N$.
\be
\frac{\partial V_N}{\partial D_{j}} = \eb^{\Tb} \left(\frac{\partial {\cal W}^{\Tb}}{\partial D_{j}} \Cb_{\Nb} {\cal W} +  {\cal W}^{\Tb} \Cb_{\Nb} \frac{\partial {\cal W}}{\partial D_{j}}\right) \eb = 0.   
\ee
The derivative of the weight matrix ${\cal W}$ can be written as:
\be
 \frac{\partial^k {\cal W}}{\partial D_{j}^k} = D_{j}^{-k} \left[ \left({\cal W} {\cal F}^{\Tb} - {\cal{I}} \right)  \Ub {\cal{J}} \Ub^{\Tb} \right]^{k} {\cal W},
\ee
with ${\cal{J}} = \frac{\partial \Db}{\partial D_{j}}$.\\

\subsection{Summary of the main MILCA steps}
 
The main steps of MILCA are the following:
(1) Filter the original data to ensure space and frequency localisation.\\
(2) Use the $N_{\mathrm{rc}}+1$ constraints on the spectral emission laws of the known components.\\
(3) Subtract the instrumental noise contribution to the covariance matrix.\\
(4) Minimize of the variance of $S_{\mathrm{1}}$ in the sub-space associated to the $N_{\mathrm{s}}-N_{\mathrm{rc}}-1$ non-constrained astrophysical components.\\
(5) Use the extra $N_{\mathrm{obs}}-N_{\mathrm{s}}$ degrees of freedom to reduce the instrumental noise contribution.\\

The final MILCA estimate of the wanted component reads
\be   
\widehat S_{\mathrm{c}} = \eb^{\Tb} \left( { {\cal F}^{T}  {\tilde \Cb}_{\Tb}^{-1} {\cal F} } \right)^{-1} {\cal F}^{T}  {\tilde \Cb}_{\Tb}^{-1} \Tb.
\ee  
For the case $N_{\mathrm{rc}}+1 = N_{\mathrm{s}}$, MILCA provides the same result as a maximum likelihood approach
\begin{align}  
\widehat \Sb = ({\Ab}^{\Tb}  \Cb_{\Nb}^{-1} \Ab)^{-1} \Ab^{\Tb}  {\Cb}_{\Nb}^{-1} \Tb.
\label{maxlike}  
\end{align}  
The $\wb$ vector cannot be constrained using the minimization of the variance of components. The last $N_{\mathrm{obs}}-N_{\mathrm{s}}$ degrees of freedom are constrained by the minimization of the instrumental noise, as it is the case for a maximum likelihood method.\\

\section{tSZ reconstruction with MILCA}
\label{reconsz}
We present here an application of the MILCA algorithm to the reconstruction of the tSZ effect. We first focus on
the two main MILCA improvements to the ILC algorithm presented in Sect.~\ref{secmod}. Then we consider the extraction
of a full-sky tSZ map using the Planck full-sky simulations presented in Sect.~\ref{secsim}.  
We consider two constrained components tSZ and CMB. Two degrees of freedom are used to minimize
the variance of the unconstrained components and the remaining two are used to minimize the
variance of the noise. The extraction of the tSZ effect has been performed at an effective resolution of 10 arcmin.\\

\begin{figure}[htbp]
\begin{center}
\includegraphics[scale=0.5]{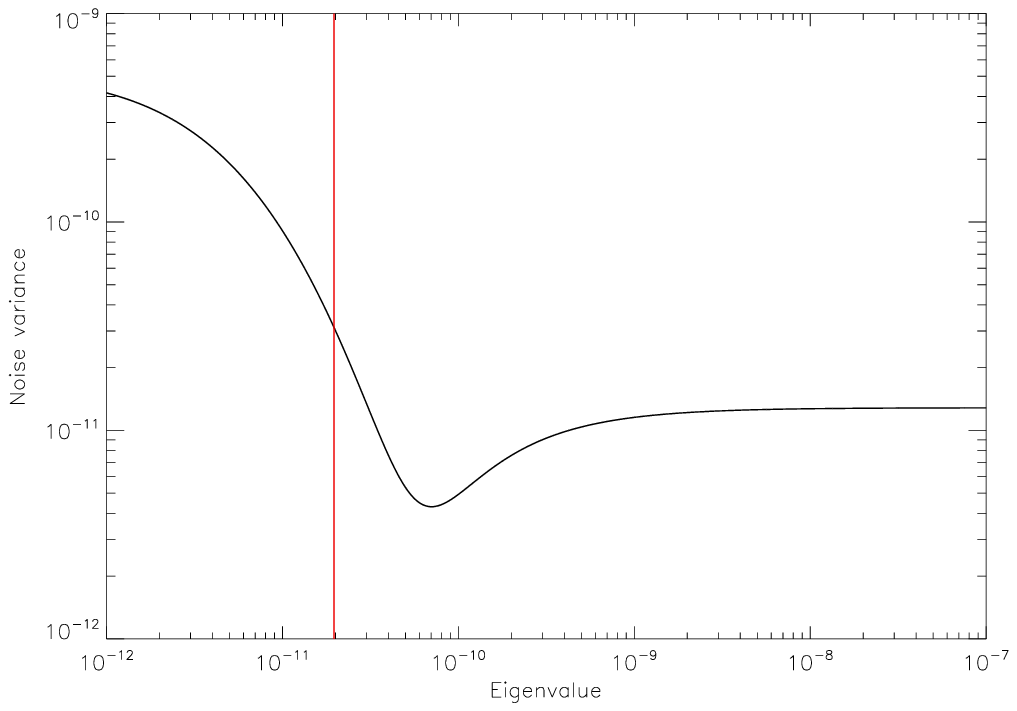}
\caption{Variation of the noise level in the reconstructed $y$-map as a function of the value of the modified eigenvalue in the data covariance matrix. The black line shows the noise level in the $y$-map, the red vertical line indicates the original value of the eigenvalue before modification.}
\label{rankred}
\end{center}
\end{figure}

\subsection{Noise variance minimization}
As discussed in Sect.~\ref{secsub} it is possible to reduce the noise contamination in the reconstructed map 
by modifying the lowest eigenvalues, $D_j$, of $\tilde{\Cb}_{\Tb}$ to minimize $V_N$.
For illustration purposes we concentrate here in the lowest eigenvalue, $D_j$, of $\tilde{\Cb}_{\Tb}$ on $V_N$, that we vary over five orders of magnitudes with respect to its original value.
For each of these values we compute the noise level in the reconstructed tSZ $y$-map that is shown as a solid
black line in Fig.~\ref{rankred}.
In this case the original value for the lowest eigenvalue of ${\Cb}_{\Tb}$ (vertical red line in Fig.\ref{rankred}) 
leads to ten times larger noise level than the optimal solution that minimize the noise variance.
Furthermore, for low values of $D_j$ an increase of the level of noise is observed as in such situation ${\Cb}_{\Tb}$ is almost singular. This produces large absolute values for $\wb$, and thus a large noise level. For large $D_j$, we converge to a rank-reduction for ${\Cb}_{\Tb}$ ($D_j$ approaches $\infty$), and this solution leads to a level of noise two times higher than the optimal MILCA solution.

\subsection{Minimization of the clustered CIB contamination using prior information}
\label{seccib}
We discuss now how we can use MILCA to reduce the clustered CIB contamination in the reconstructed tSZ map.
Reducing clustered CIB contamination is a major issue for tSZ analysis, because the CIB is, on the one hand, correlated with the tSZ emission \citep{add12} and on the other hand, it is a diffuse emission that cannot be masked in tSZ maps.\\
The clustered CIB component cannot be described with a single template and an SED; furthermore, it is only partially correlated from a frequency to another \citep{planck11b}. 
Indeed, the CIB is produced by the infra-red emission from extra-galactic sources at different redshifts and their observed emission law is highly dependent on the redshift. Consequently, different frequency bands are not sensitive to the same sources, according to their redshifts.\\
We thus cannot apply constraints on the SED, in the ${\cal F}$ matrix, in order to reduce the clustered CIB contamination. 
By contrast we can use the approach proposed in Sect.~\ref{subsect:p2c1:tilt} by modifying the data covariance matrix.
The covariance matrix of the clustered CIB emission needs to account for the CIB SED and for partial correlation between frequencies and it can be inferred from models or from the data themselves.\\

\begin{figure}[htbp]
\begin{center}
\includegraphics[scale=0.5]{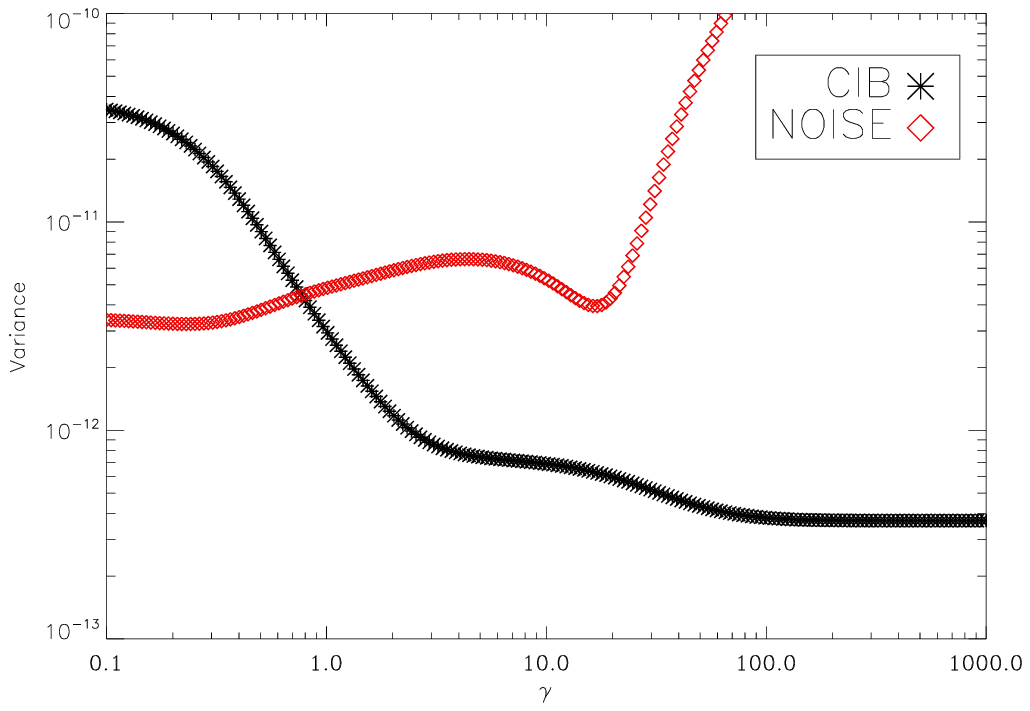}
\caption{Evolution of the clustered CIB contamination (black stars) and the noise level (red ) as a function of $\gamma$. Both contributions (clustered CIB and noise) are presented in units of variance in the reconstructed tSZ $y$-map.}
\label{cibconta}
\end{center}
\end{figure}

In Fig.~\ref{cibconta}, we present the variation of the clustered CIB contamination and the reconstructed tSZ map noise level as a function of 
the parameter $\gamma$ in equation~\ref{modifcovmatrix}. $\gamma <1$ corresponds to the case where we do not account for the clustered CIB
and it leads to large clustered CIB contamination. As expected, the CIB contamination decreases with 
as $\gamma$ increases. However, as the clustered CIB contamination decreases the noise level increases.
A compromise between noise and clustered CIB contamination can be found for $\gamma$ in the range between 3 and 30, as we observe that the noise level is almost constant in this range and the clustered CIB contamination decreases with a slope $\simeq \gamma^2$. Notice that for $\gamma=1$ the
clustered CIB contamination is already reduced by one order of magnitude in variance with respect to the standard ILC algorithm.
On the real data such a compromise can also be found using estimates of both the noise and clustered CIB covariance matrices.
At large values of $\gamma$, we observe that the clustered CIB contamination becomes constant. This behavior is produced by the partial correlation of the clusterd
CIB across frequencies. Indeed, as the CIB emission is not fully correlated between frequencies, it can not be totally removed.\\

\subsection{Characterization of noise level and bias on full-sky simulations}

\begin{figure}[htbp]
\begin{center}
\includegraphics[scale=0.25]{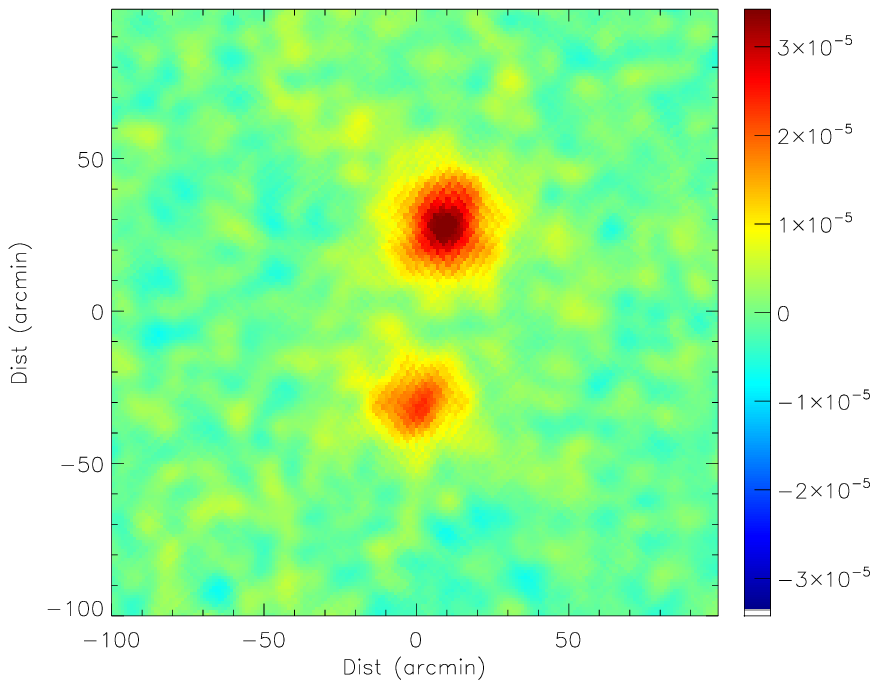}
\includegraphics[scale=0.25]{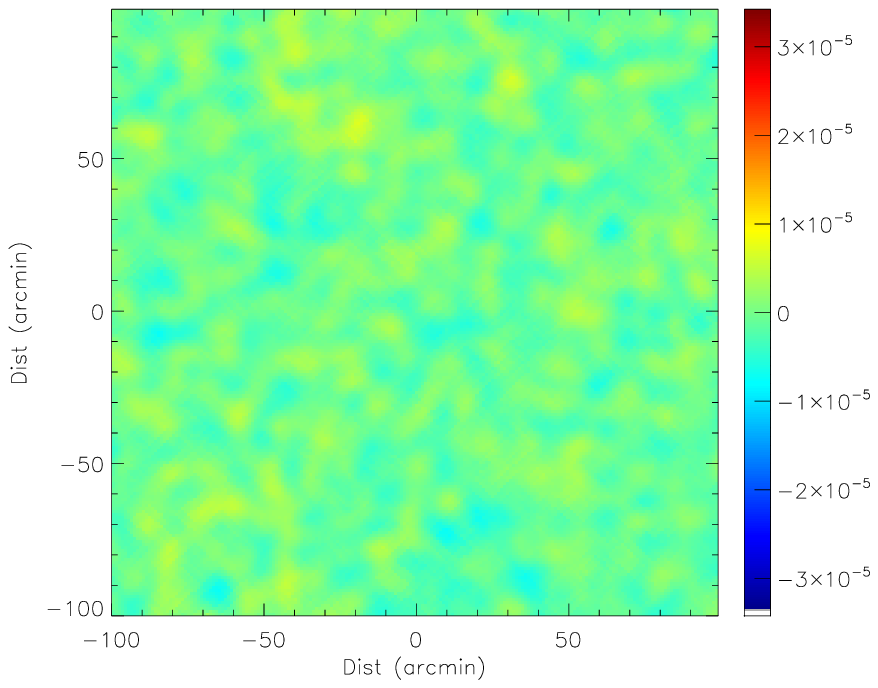}
\caption{Left panel: reconstruction of a tSZ $y$-maps from simulated Planck data using the MILCA approach. Right panel:  residual map after subtraction of the input tSZ signal.}
\label{a399}
\end{center}
\end{figure}

We present in Fig.~\ref{a399}  the reconstructed tSZ map and the residual map in a small patch of 200 by 200 arcmin centered on the well-known pair of clusters A399-A401 \citep{planck2012-VIII}. This example illustrates the ability of MILCA to properly deal with complex tSZ systems like mergers. \\

\begin{figure}[htbp]
\begin{center}
\includegraphics[scale=0.5]{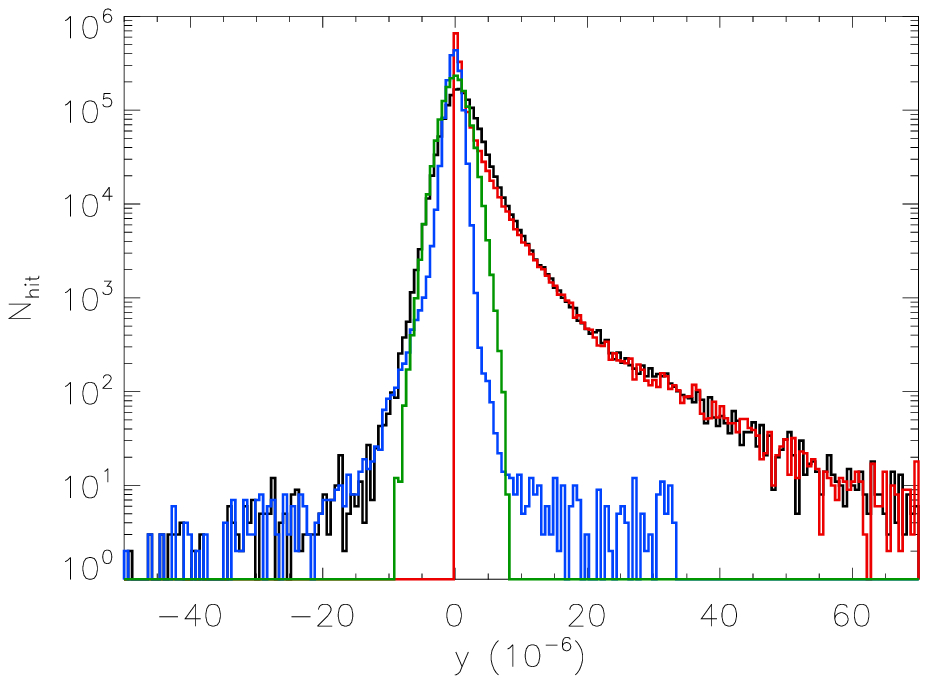}\\
\includegraphics[scale=0.5]{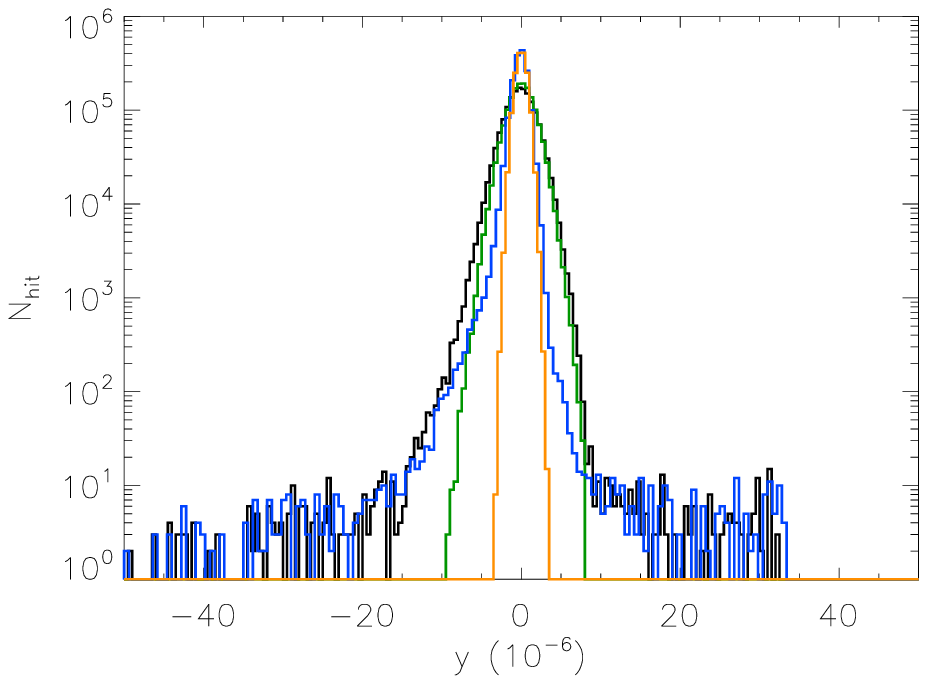}
\caption{Top: 1D pdf for the MILCA tSZ map (black), for the tSZ input map (red), for the noise (green) and the contamination by other components (in blue) . Bottom:  1D pdf for the residuals emission (black), for the noise (red), for the contamination by all other components (blue) and for the CIB component contribution (green).}
\label{1pdf}
\end{center}
\end{figure}

In a more quantitative way we present in Fig.~\ref{1pdf} the 1D probability density function (1D pdf) of the reconstructed
MILCA tSZ map for all pixels located inside a radius of 15 arcmin from any of the 1743 clusters of the MCXC catalog \cite{pif10}. 
In the top panel we present the main contributions to the reconstructed tSZ map: tSZ emission (black), noise (green), and other astrophysical components
(blue). For comparison we also plot the input tSZ signal (red). We observe that the reconstructed tSZ map is dominated by the tSZ signal.
In the bottom panel we represent the 1D pdf of the residual map (same color scheme as before).
The residual map is dominated by instrumental noise which contributes to the residual rms at a level of $1.7 \times 10^{-6}$ (Compton parameter units),
then, the CIB contributes to a level of $0.7 \times 10^{-6}$ and, all other components have a total contribution of about $1.0  \times 10^{-6}$.\\ 
Notice in Fig.~\ref{1pdf} that most of the residuals by other astrophysical components (in blue) appear as a negative bias, they are mainly produced by radio sources.
An extreme case of such a bias is discussed in Sect.~\ref{secper}. We also observe small contamination by IR point sources which appear as positive bias in the reconstructed map. Other astrophysical emissions marginally contribute to the reconstructed signal.\\
An example of the application of MILCA to real Planck data is presented in Sect.~\ref{secmilreal}.


\subsection{Empirical extra-constraints for tSZ extraction}
\label{secper}

We present an application of MILCA to an extreme case of radio-loud AGN contamination for a simulated Perseus-like cluster. 
This particular cluster is well known to be very extended over the sky and to host a radio-loud AGN  \citep{bru99}. 
The AGN emission makes the extraction of the tSZ signal for such kind of systems quite complex. However, as the tSZ signal is extended, it is possible to separate the two emissions (from the cluster and from the AGN).\\

\begin{figure}[htbp]
\begin{center}
\includegraphics[scale=0.25]{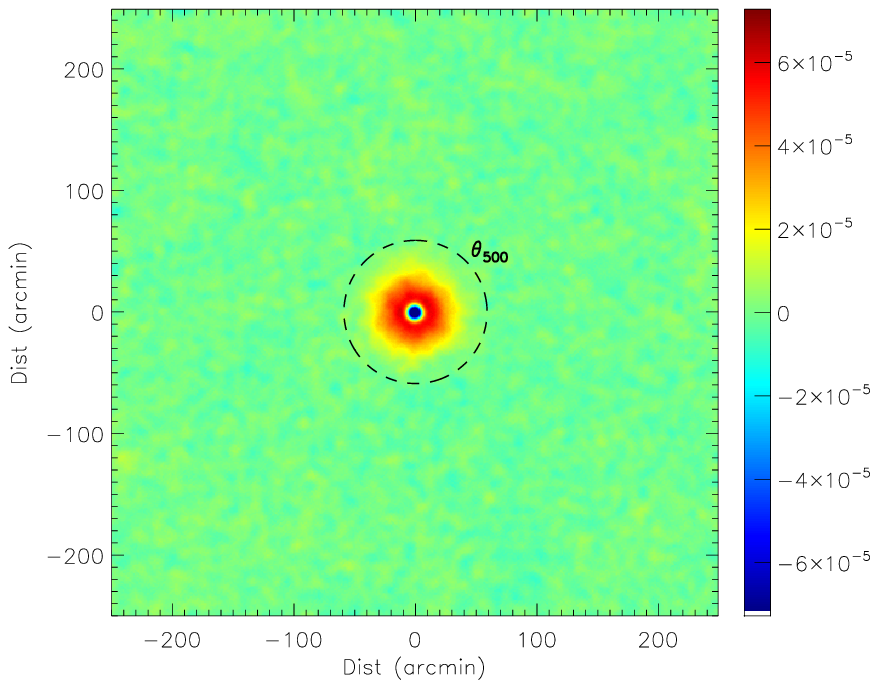}
\includegraphics[scale=0.25]{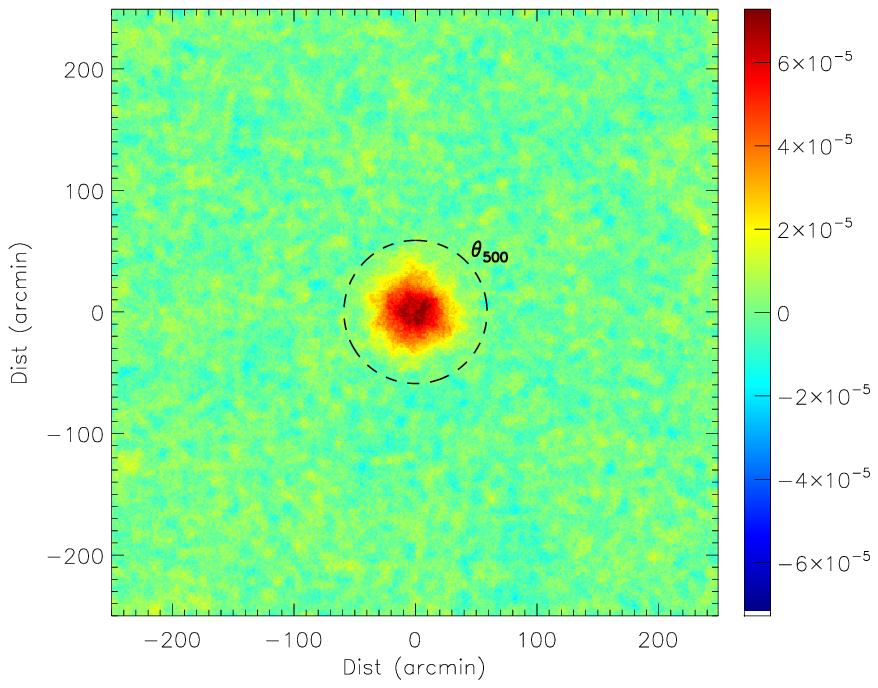}\\
\includegraphics[scale=0.25]{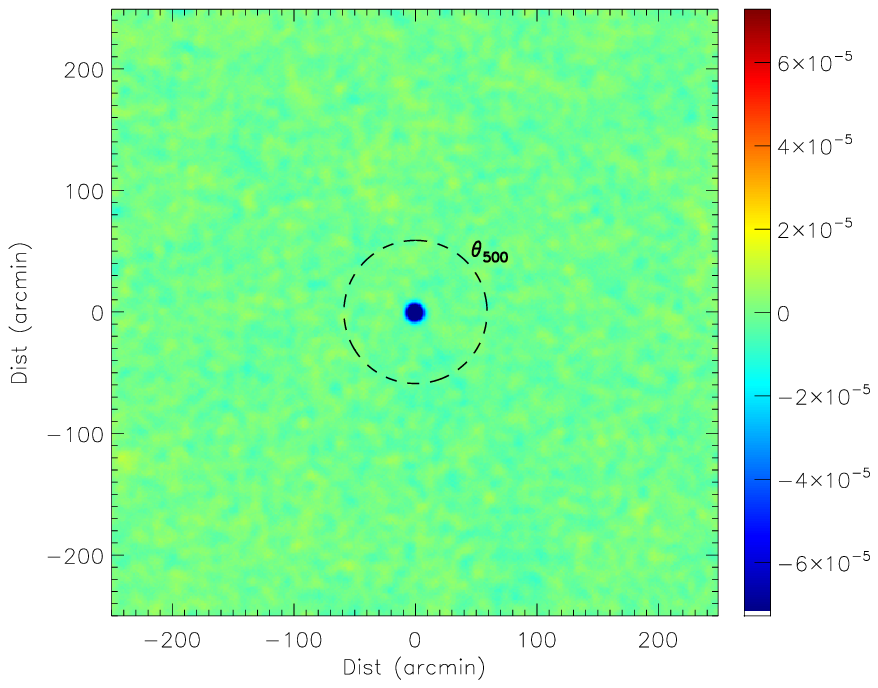}
\includegraphics[scale=0.25]{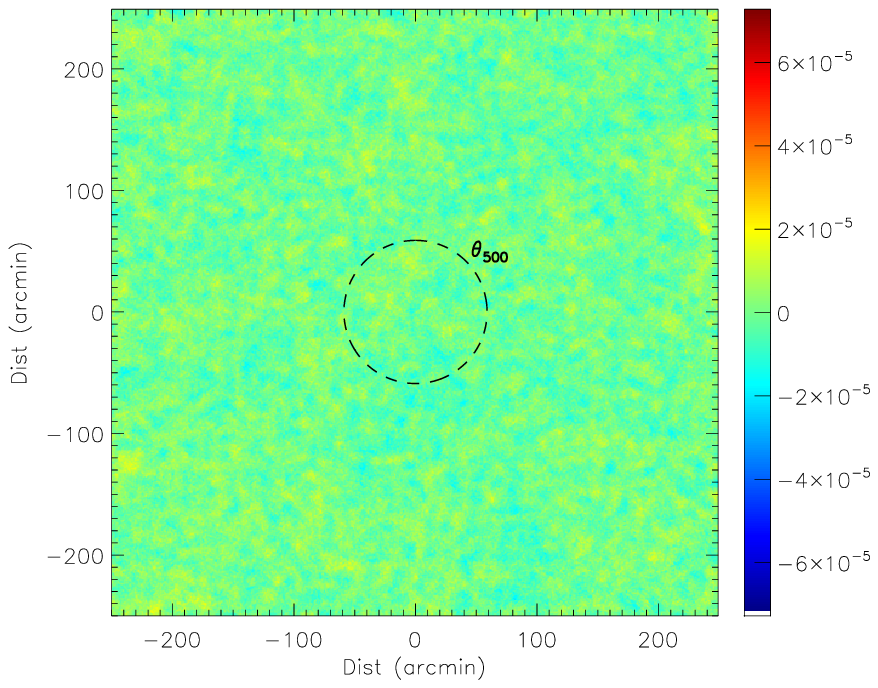}
\caption{Reconstructed $y$-map (top) and residuals (bottom) for the Planck simulated data using a standard ILC adapted to the tSZ effect (left panel) and
using MILCA with three constrained components (right panel).
\label{perseus3const}}
\end{center}
\end{figure}

Figure~\ref{perseus3const} presents the comparison between the tSZ $y$-maps (top) and residuals obtained from a standard ILC adapted to the tSZ effect (left panel)  and MILCA (right panel) using three constraints to keep the tSZ signal but removing the CMB and the central radio-loud AGN contamination. 
We use an estimate of the AGN SED obtained directly from the data themselves. 
For the standard ICL case we observe a strong contamination by the AGN emission (with an amplitude of $-9 \times 10^{-5}$ in Compton parameter units at the center
of the AGN). In this case the tSZ effect and the AGN radio emission are strongly spatially correlated over the sky, leading to the bias  discussed in Sect.~\ref{secbias}, when a correlated component is used to clean the tSZ effect contribution. 
When considering an extra constraint, we are able to recover the tSZ signal with a much smaller bias from the AGN (an amplitude of about $-0.5 \times 10^{-6}$). This extra constraint presents however some drawbacks, such as an increase of the noise level (in this case by a factor of 1.5 ) and/or larger bias from other astrophysical un-constrained components.\\

The ability of MILCA to reduce such contamination is directly related to the accuracy of the estimated SED used for the extra constraint. In order to quantify this we performed simulations of the reconstruction but with an extra-constraint applied on a noisy SED for the AGN
\be
f_i = f^{\mathrm{AGN}}(1+\sigma X_i),
\ee
where $f^{\mathrm{AGN}}$ is the AGN simulated SED, $X_i$ is a random variable following a standard normal distribution and $\sigma$ is the relative error on the AGN SED.
We performed 200 simulations of reconstruction for a grid of 100 values of $\sigma$. 
\begin{figure}[htbp]
\begin{center}
\includegraphics[scale=0.5]{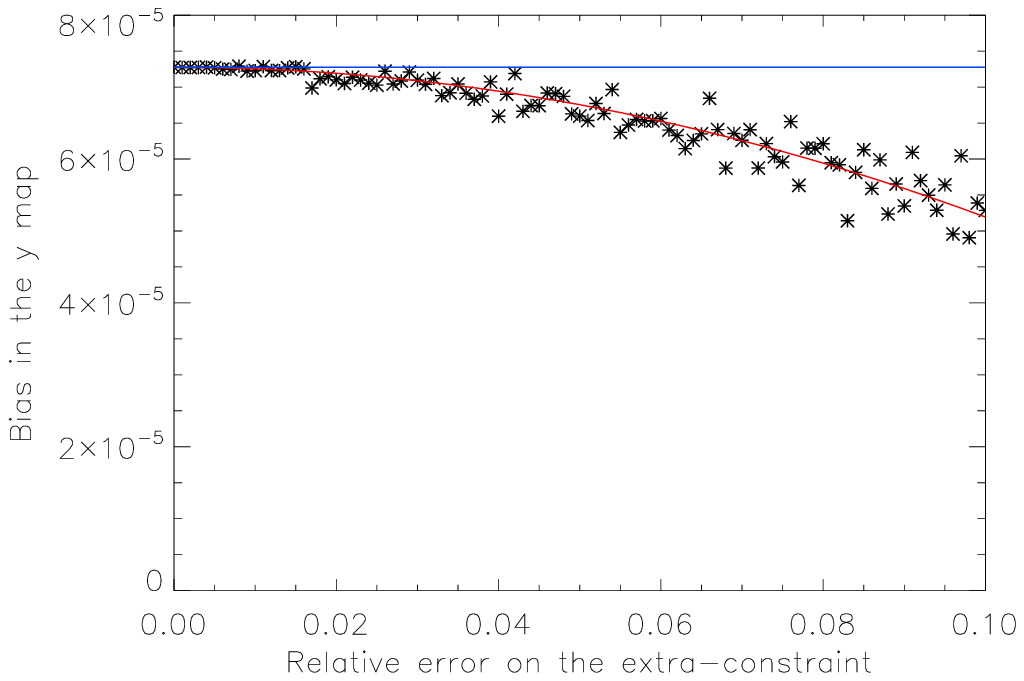}
\includegraphics[scale=0.5]{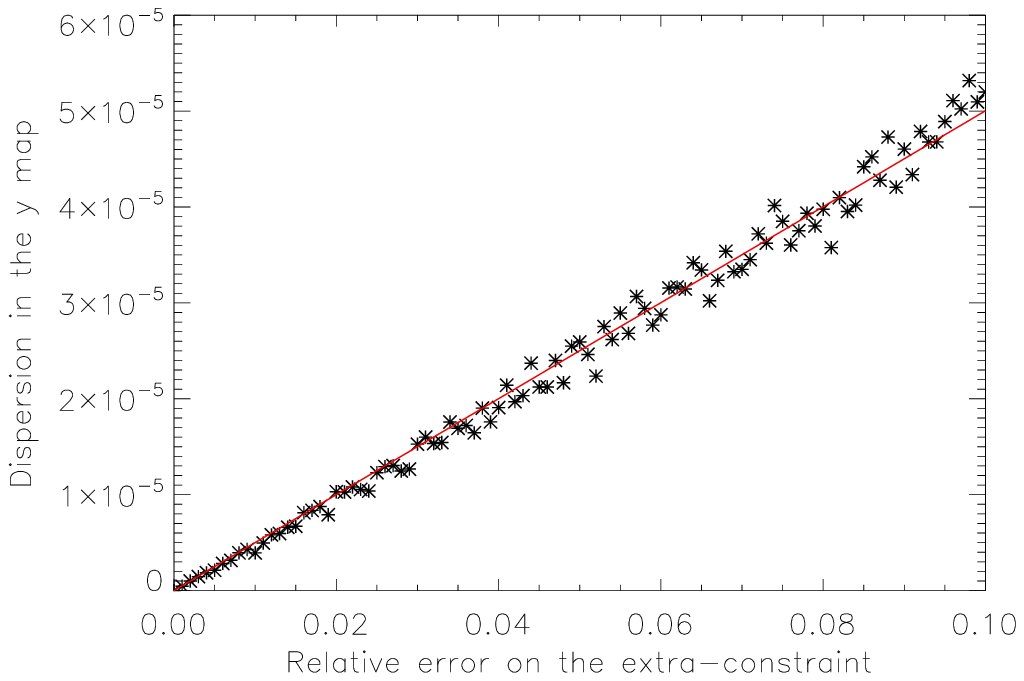}
\caption{Top panel : bias induced in the MILCA reconstruction when using a noisy SED as extra-constraint. Black points are the mean from the 200 simulations for each value of the relative error, the red line represents the best-fit for the bias term and the blue line is the expected value in an unbiased case. Bottom panel : extra noise induced in the MILCA reconstruction when using a noisy SED as extra-constraint. Black points are computed from the 200 simulation for each value of the relative error and the red line represents the best-fit for the noise term. The contamination are presented in units of Compton parameter for the central pixel of the reconstructed map for which we expect the maximum contamination.}
\label{biasnoise}
\end{center}
\end{figure}
A global bias is computed by averaging the 200 reconstructions and the noise rms is obtained from the standard deviation of the 200 reconstructions. 
Figure~\ref{biasnoise} shows the bias and the extra noise induced on the reconstruction by a noisy SED. 
The bias is proportional to the square of the relative error, $\sigma$, while the extra-noise is directly proportional to $\sigma$. 
This relation allows us to propagate both bias and noise from the used SED to the final reconstructed map. 
Notice that the calibration of this relation depends on both the AGN amplitude and the local background of the cluster. Consequently, such estimation of the noise and bias induced by a noisy SED have to be computed for each specific case.

\begin{figure}[h!]
\center
\begin{tabular}{p{2.365cm}p{2.365cm}p{2.365cm}}
~~~~~~~~MILCA & ~~~~~~~~~~ILC & ~~~~~~~Residual\\
\includegraphics[width=4cm]{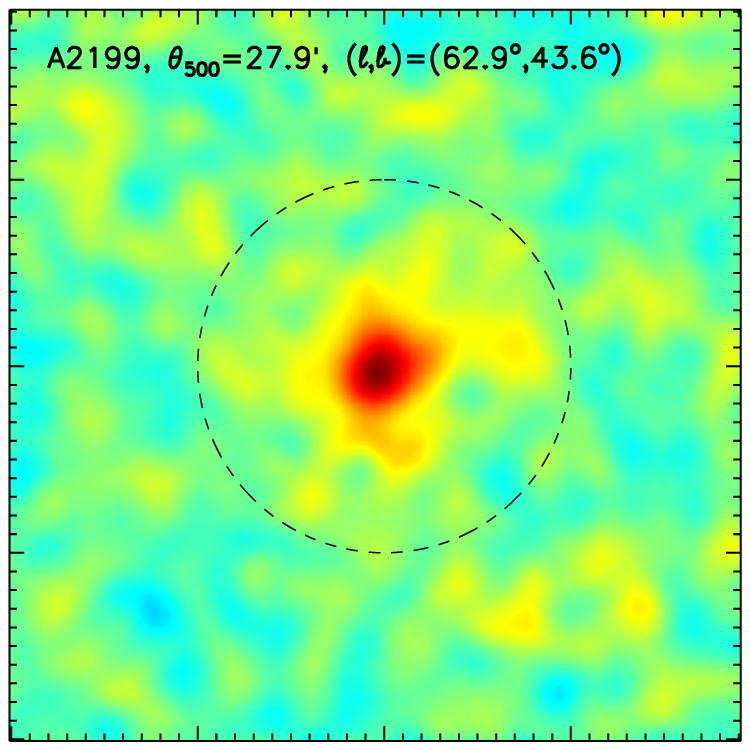} &
\includegraphics[width=4cm]{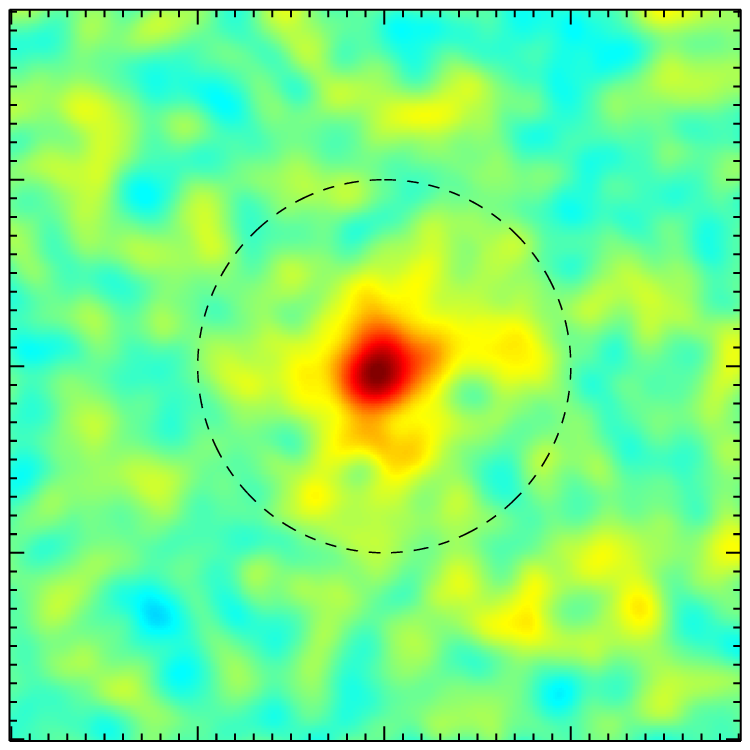} &
\includegraphics[width=4cm]{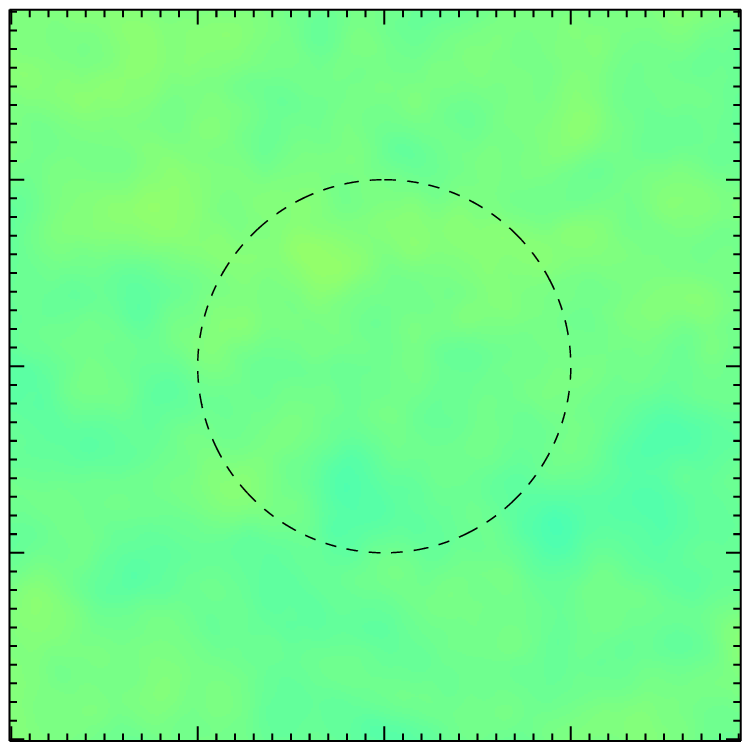} \\[-0.24cm]
\includegraphics[width=4cm]{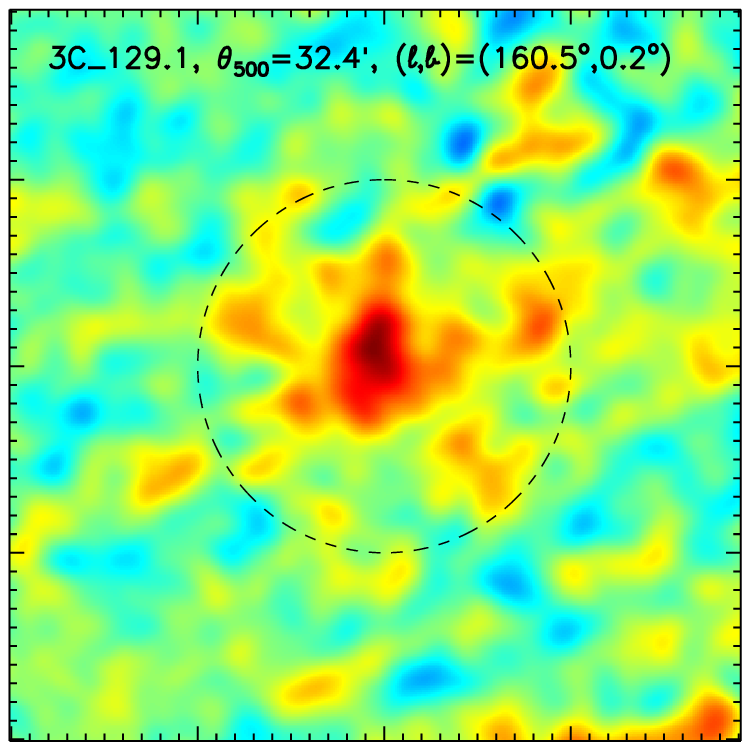} &
\includegraphics[width=4cm]{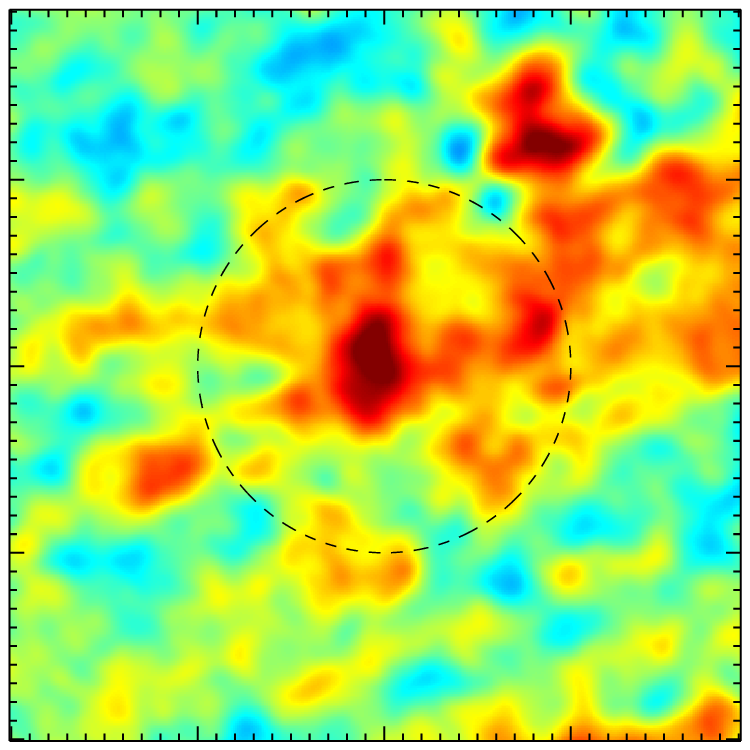} &
\includegraphics[width=4cm]{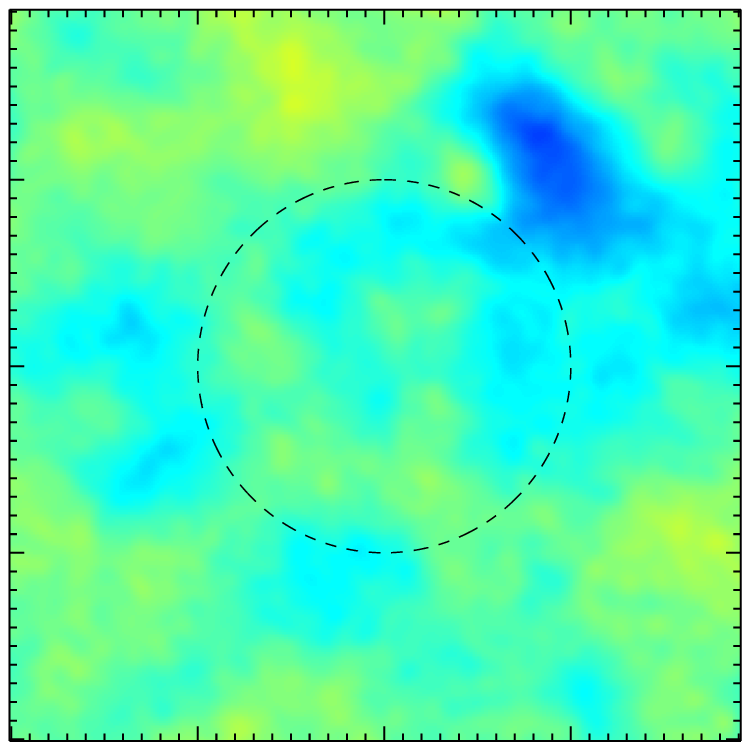} \\[-0.24cm]
\includegraphics[width=4cm]{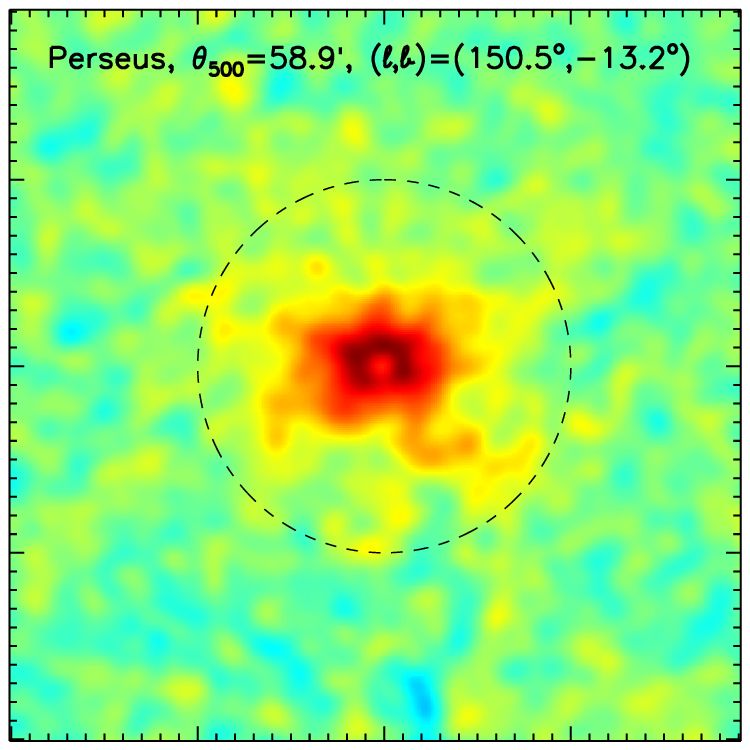} &
\includegraphics[width=4cm]{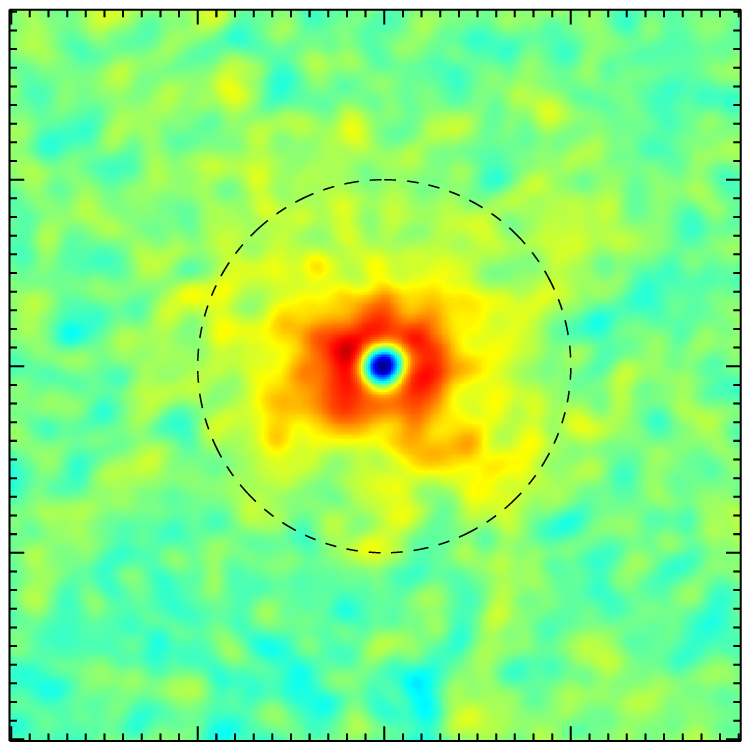} &
\includegraphics[width=4cm]{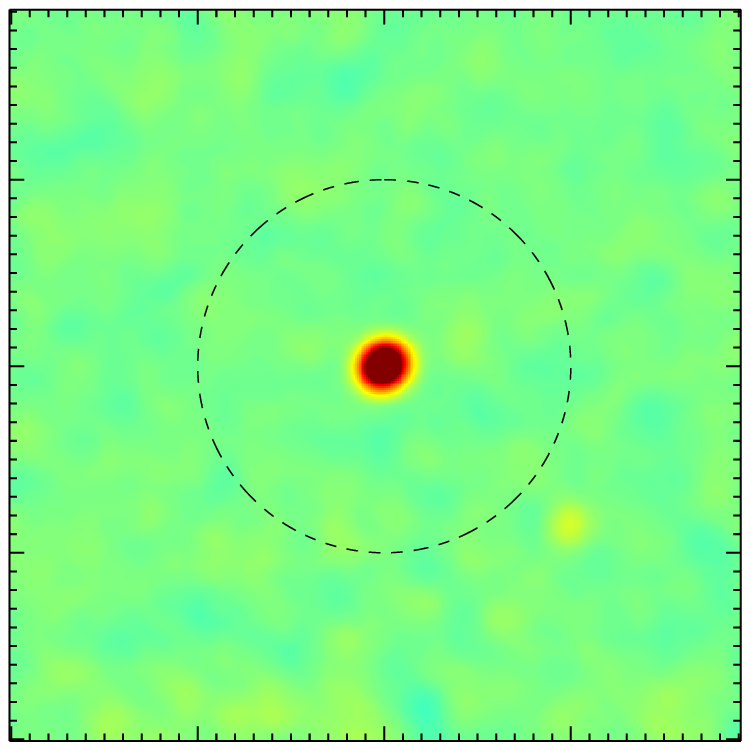} \\[-0.24cm]
\includegraphics[width=4cm]{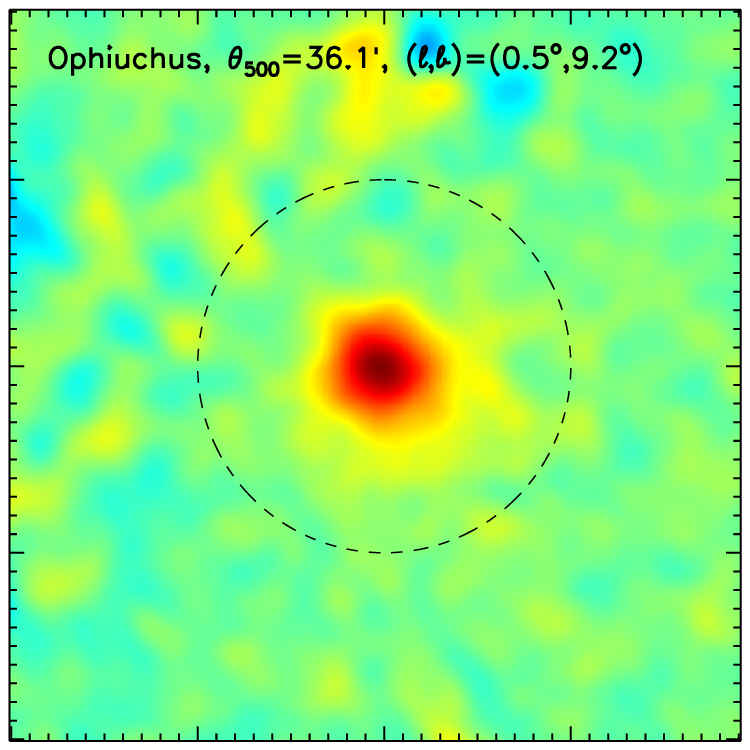} &
\includegraphics[width=4cm]{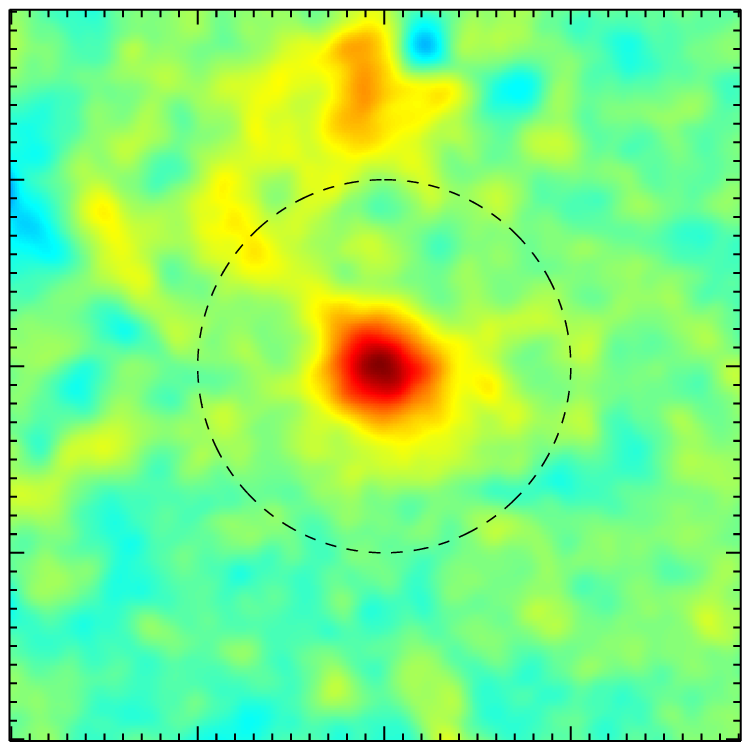} &
\includegraphics[width=4cm]{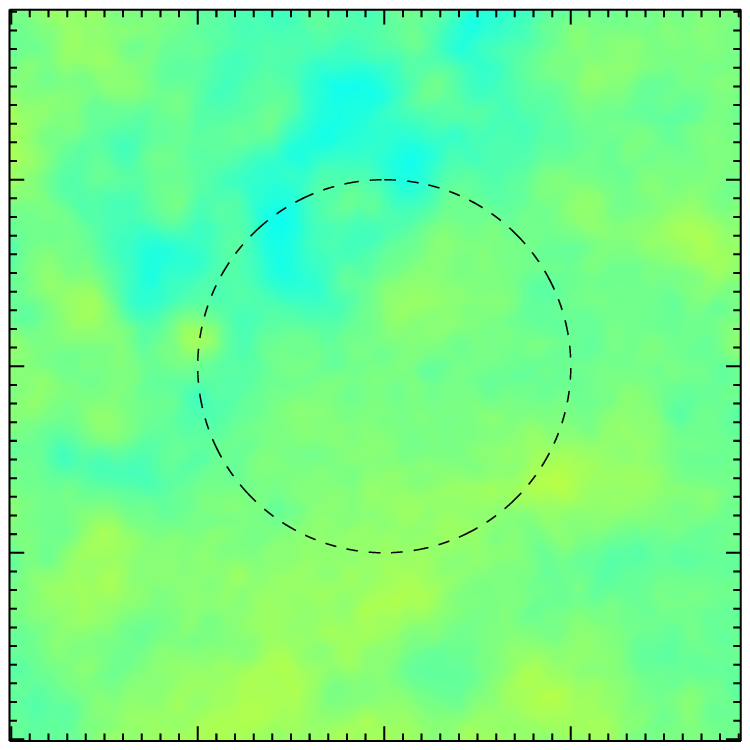} \\[-0.24cm]
\includegraphics[width=4cm]{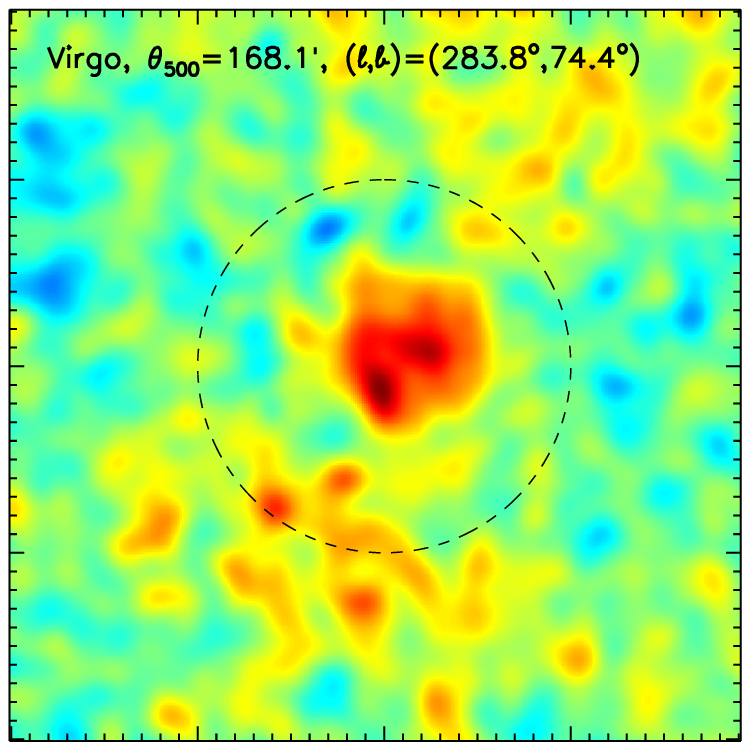} &
\includegraphics[width=4cm]{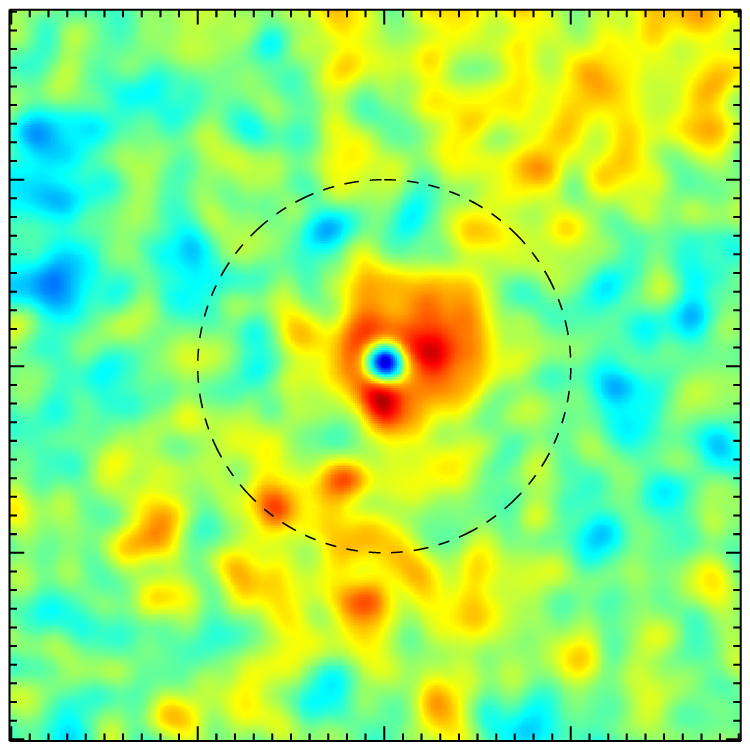} &
\includegraphics[width=4cm]{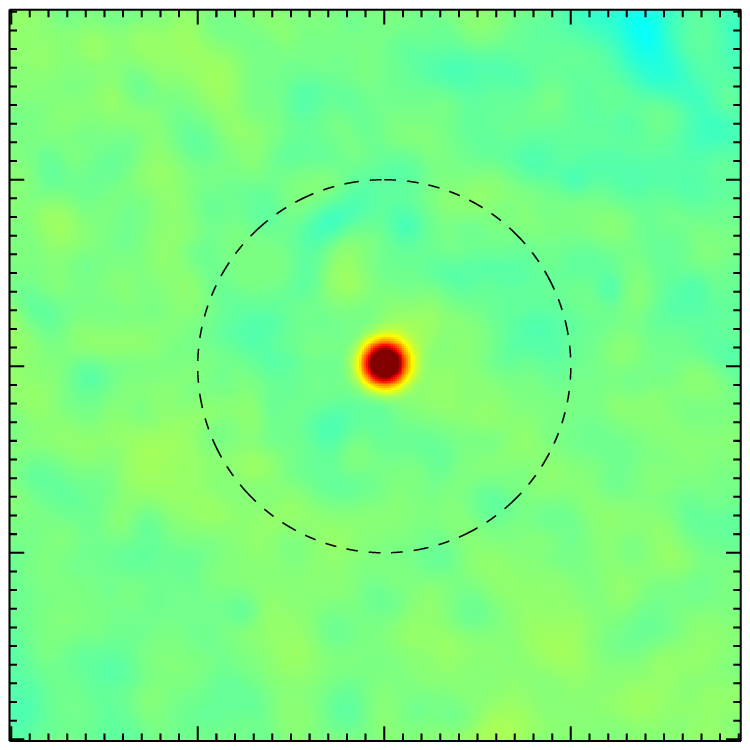} \\[-0.24cm]
\includegraphics[width=4cm]{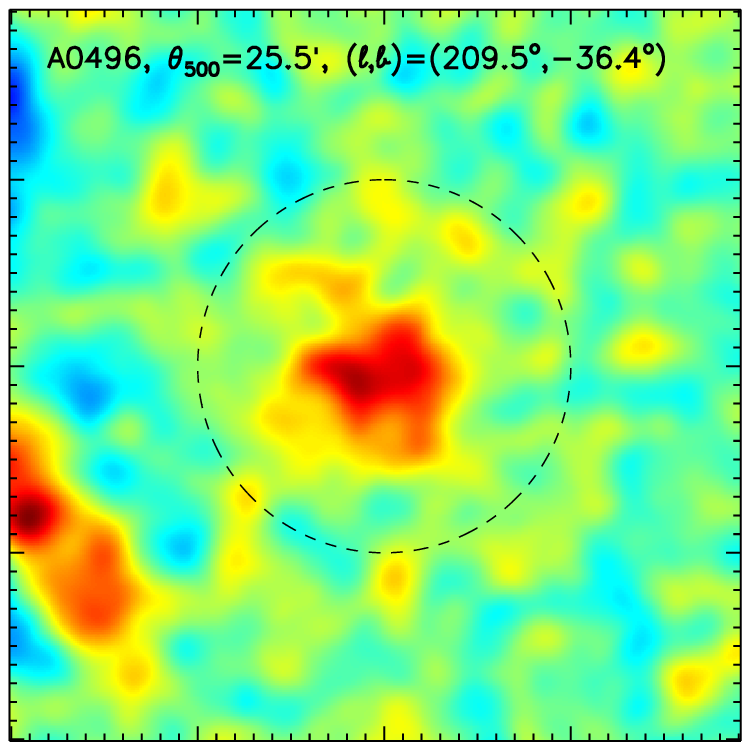} &
\includegraphics[width=4cm]{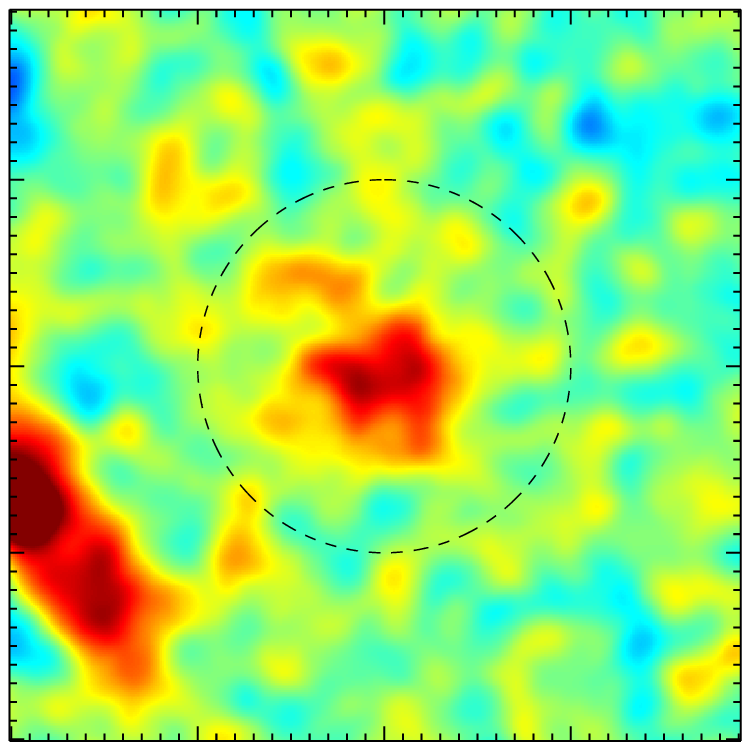} &
\includegraphics[width=4cm]{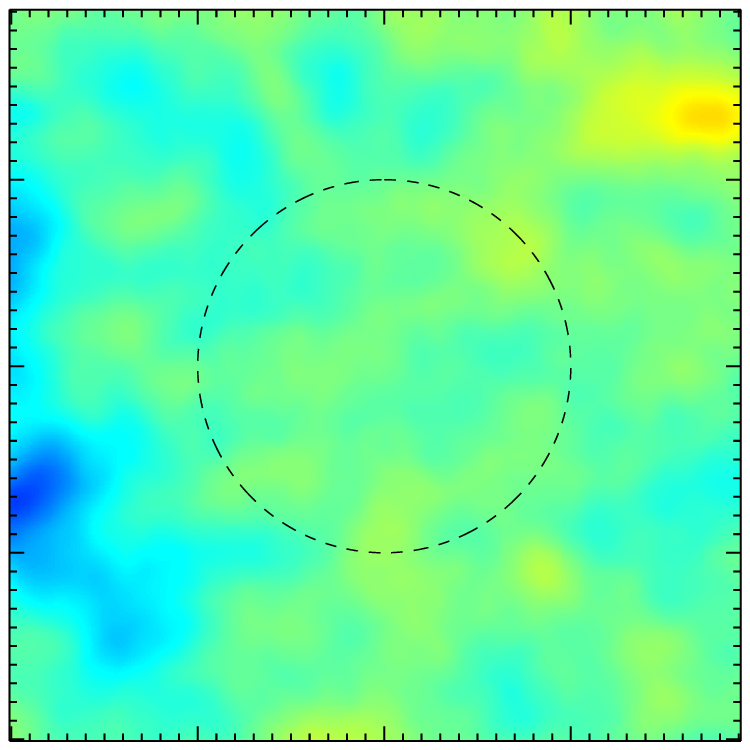} \\[-0.24cm]
\includegraphics[width=4cm]{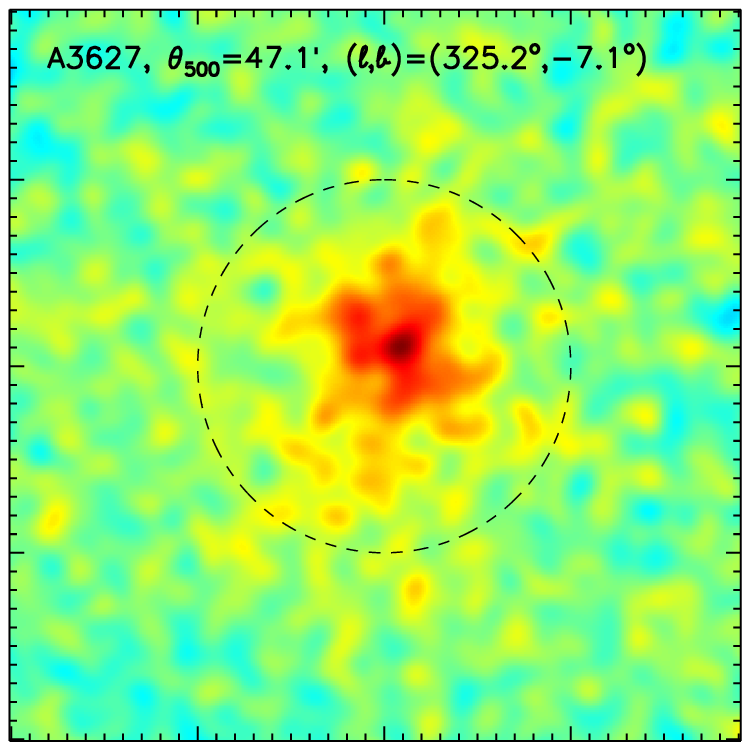} &
\includegraphics[width=4cm]{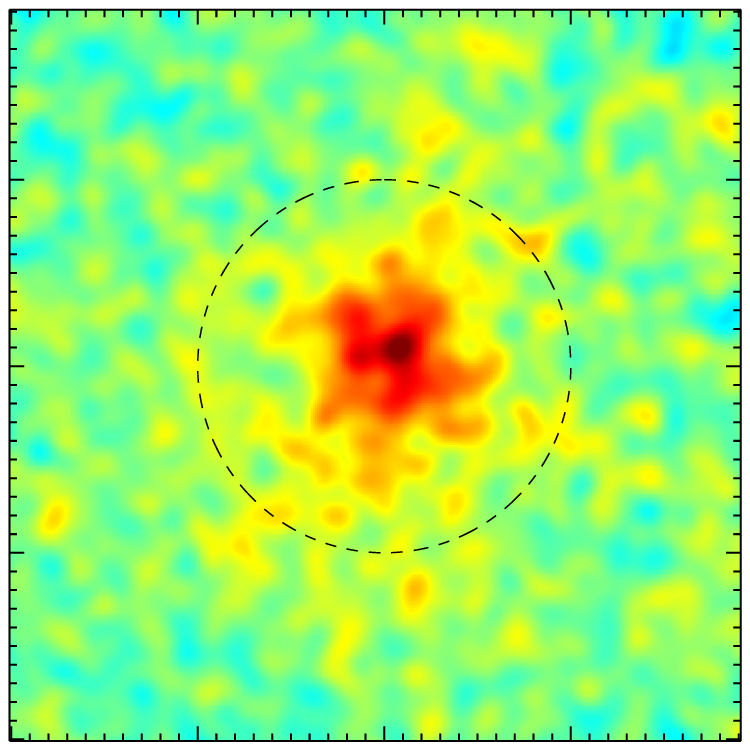} &
\includegraphics[width=4cm]{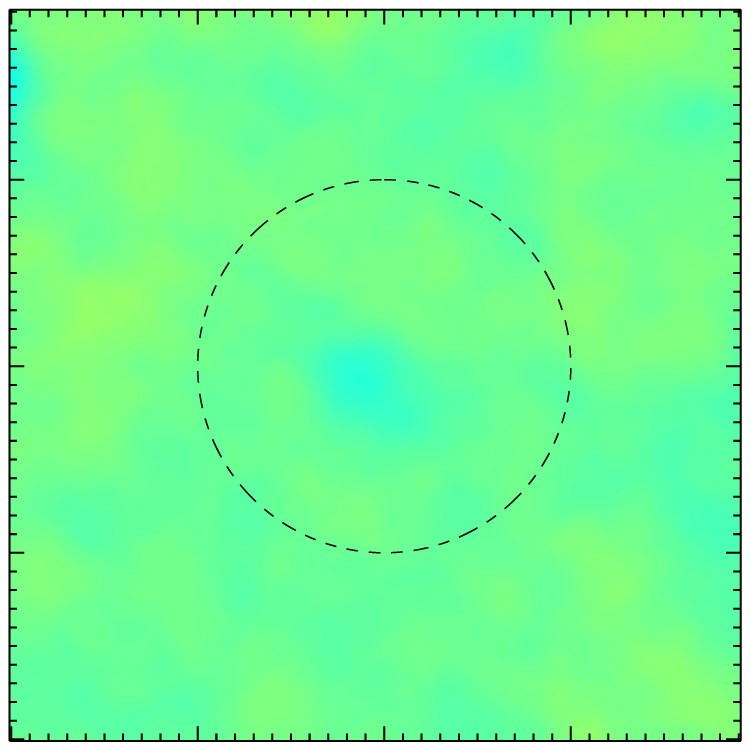} \\[-0.24cm]
\end{tabular}
\caption{From left to right: tSZ reconstructed $y$-map with MILCA, with a tSZ effect adapted standard ILC and the difference between the 2 reconstructions. 
Each row represents the reconstructions for a particular cluster in a FOV of $4~\times~R_{500}$, with the dashed circle representing the $R_{500}$ aperture. For each cluster the three maps are displayed with the same color scale. All maps are presented at a resolution of 7.18' FWHM except the Virgo maps at 30' FWHM.}
\label{milreal}
\end{figure}
\subsection{Extraction of the tSZ effect with MILCA on real data}
\label{secmilreal}
For illustration in Fig.~\ref{milreal}, we present tSZ effect reconstructed MILCA (left) maps using the Planck public data for very extended clusters over the sky  \citep[see also][]{planck2013-p05a}. We also present the comparison with a standard ILC reconstruction (middle). These figures illustrate the improvement provided by MILCA for 
the reconstruction of tSZ $y$-maps, not only on simulations, but also on real datasets.\\
For some clusters, such as A2199, Ophiucus and 13627, MILCA and a standard ILC adapted to tSZ extraction produce very similar results. Indeed, for these clusters the foreground contamination is low and not correlated with the tSZ component.\\
However in the case of AGN contaminated clusters, such as Perseus and Virgo, we observe a clear contamination in the standard ILC maps that is significantly reduced in the MILCA maps. For the Perseus galaxy cluster we observe similar results than the ones obtained on simulated datasets in Sect.~\ref{secper}. The use of an extra-constraint allows MILCA to reduce the contamination by radio-loud AGNs, which are correlated to the tSZ effect in the cluster.\\
For the 3C~129.1 cluster, we observe in the standard ILC map, a large contamination around the tSZ emission. This contamination is mainly produced by thermal dust residuals. This cluster is located in the galactic plane. Consequently, there is a large contamination by dust in this area. However, the MILCA maps are less contaminated by the thermal dust emission. This reduction of the contamination by thermal dust is  mainly due to the correction of the noise-induced bias.\\ 
Finally for A0496, we observe a reduction of the contamination in MILCA map
but we still observe contamination by an IR point source. This IR point source is very faint with respect to the thermal dust background. Consequently, it is difficult to extract the SED of this source, reducing the ability of MILCA to minimize the source contamination in the reconstructed tSZ map.\\

MILCA has been used in the context of the \textit{Planck} collaboration to produce tSZ $y$-maps for scientific exploitation such as, validation of the catalog of candidates \citep{planck11d,planck11e}, for the extraction of the cluster pressure profiles \citep{planck13b}, for other physicals analysis \citep{planck13c,planck13e} and for comparison with other experiment \citep{planck13d}. MILCA has also been used to produced tSZ maps in \citet{planck11d,planck11e} and \citet{hurin3,hurin4,hurin5,planck2013-p05a,plancksz}.\\

\begin{figure*}[!th]
\begin{center}
\includegraphics[scale=0.3,angle=90]{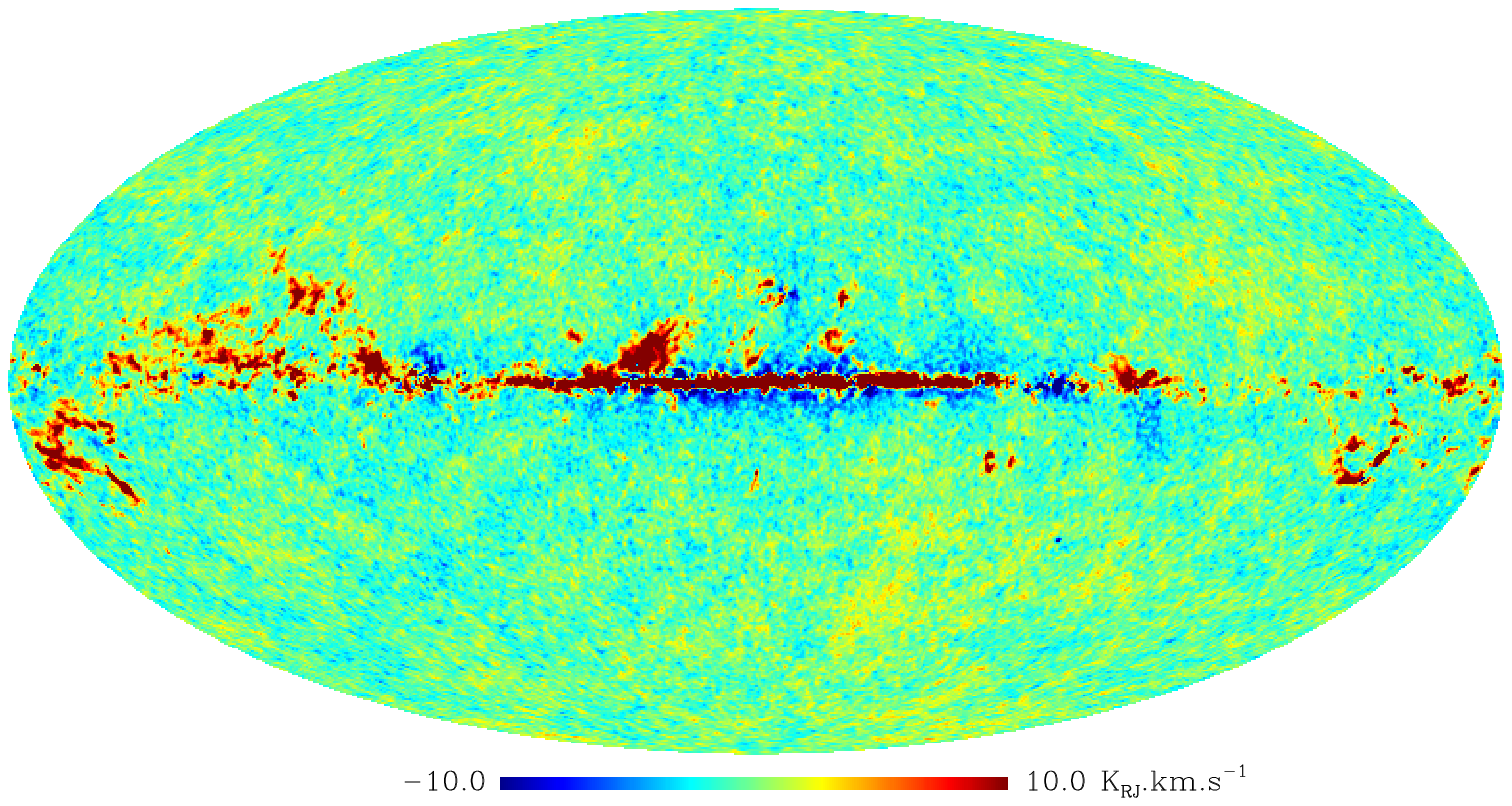}
\includegraphics[scale=0.3,angle=90]{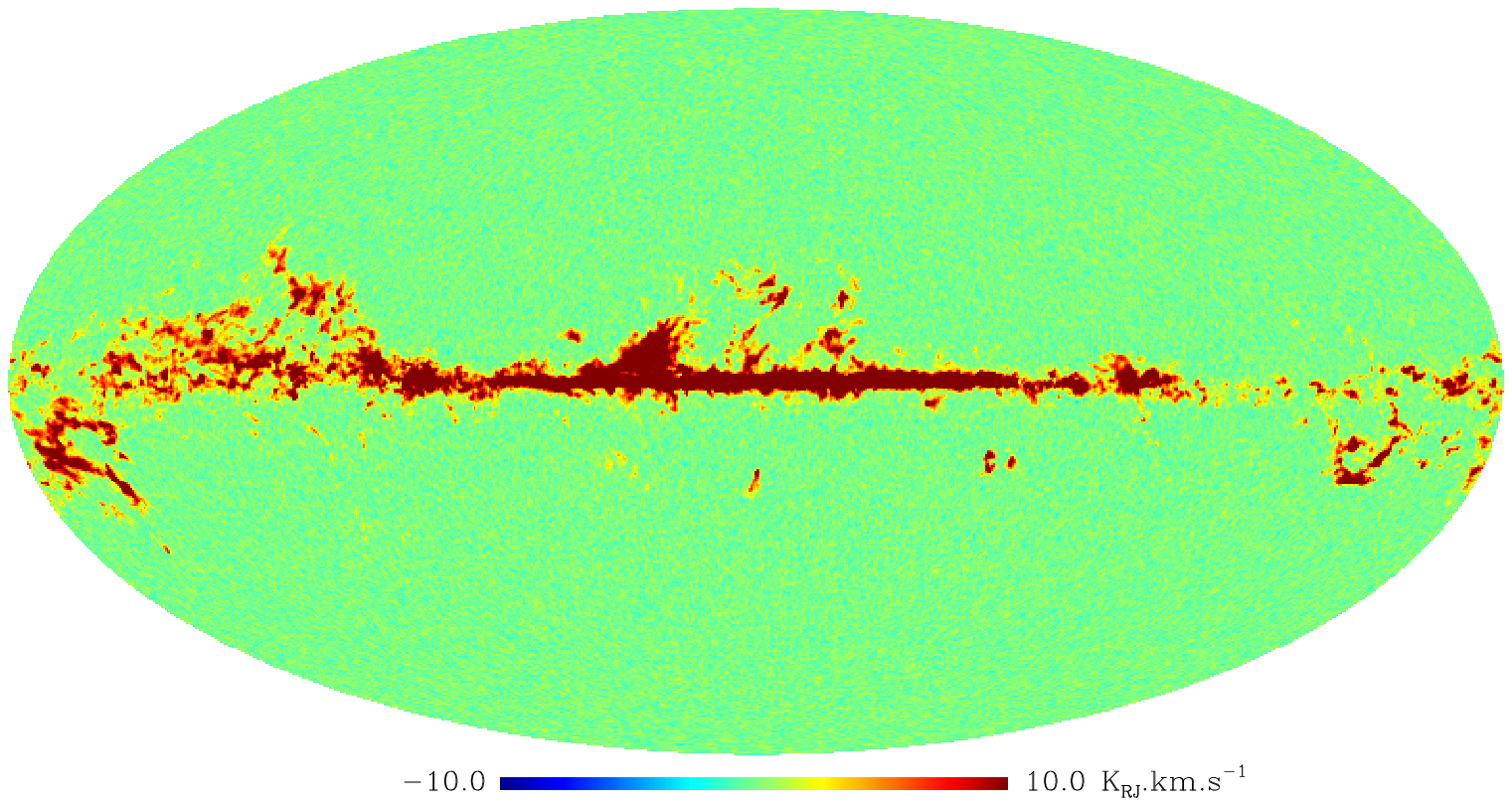}\\
\includegraphics[scale=0.3,angle=90]{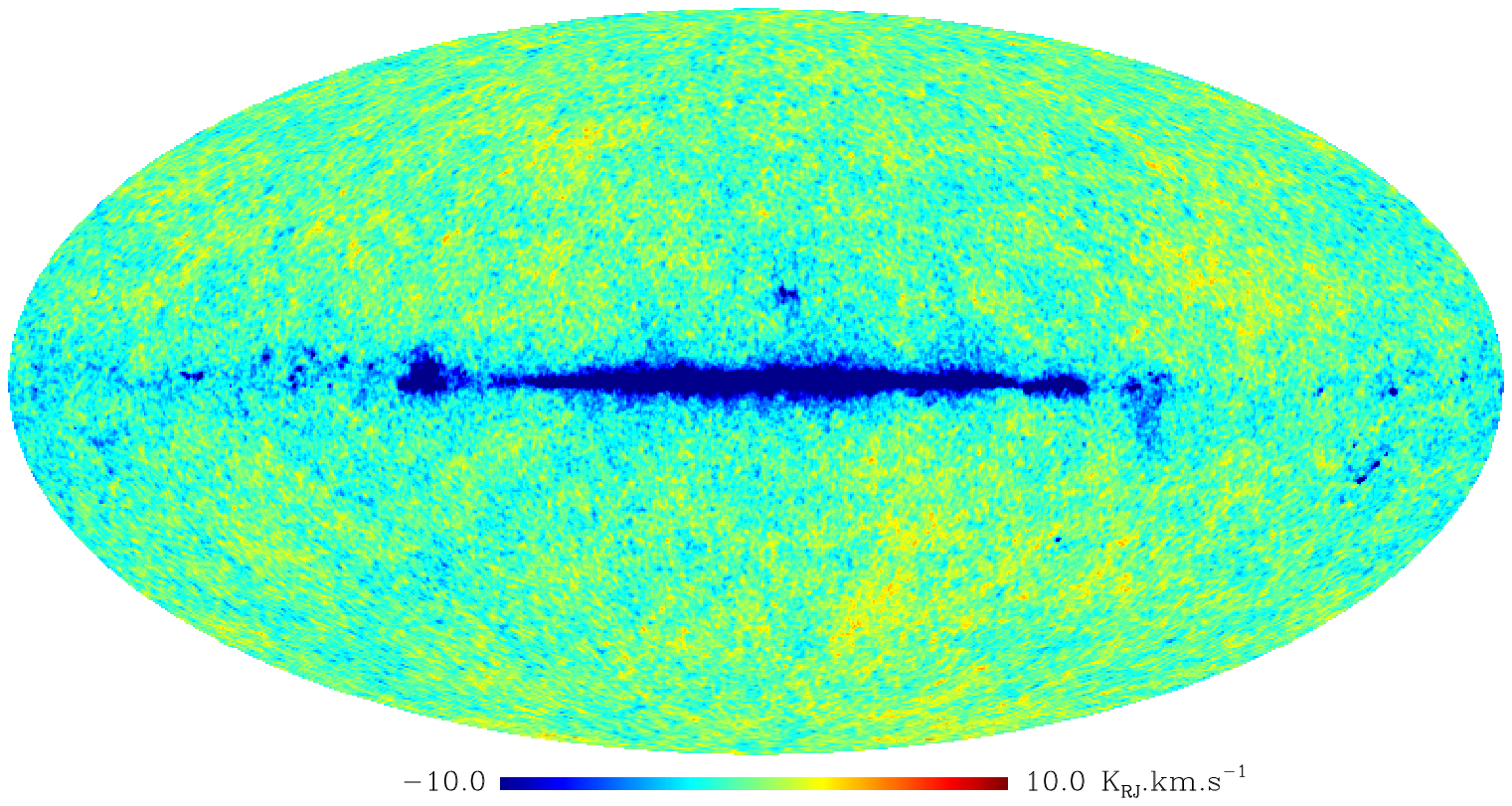}
\includegraphics[scale=0.3,angle=90]{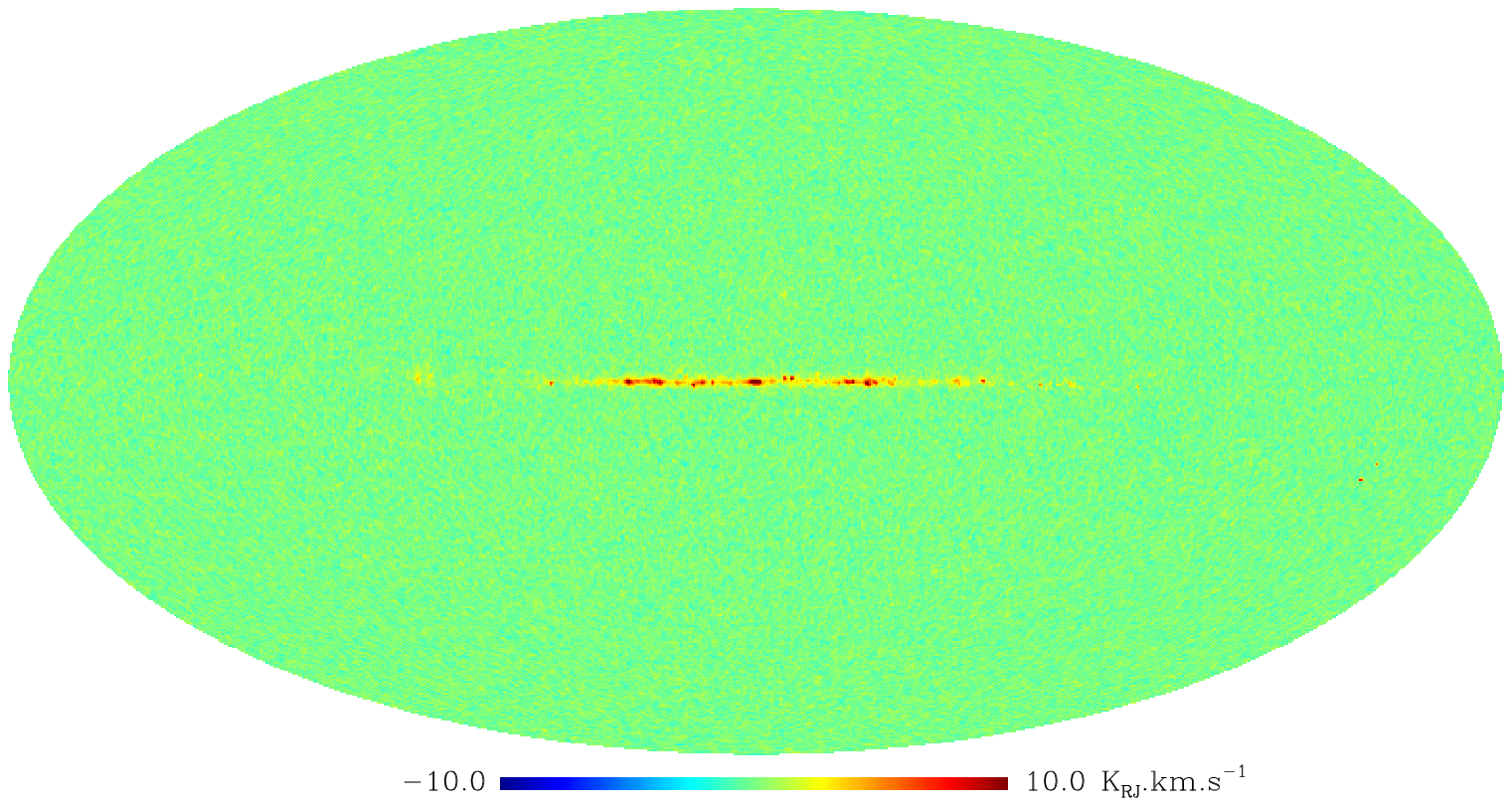}
\caption{Reconstructed CO emission map (top) and residuals (bottom) for the CO adapted algorithm (left) and for MILCA (right). The maps are presented at a resolution of 30
arcmin to reduce the noise contribution in the maps.}
\label{milcaco}
\end{center}
\end{figure*}

\begin{figure}[h!]
\begin{center}
\includegraphics[scale=0.25]{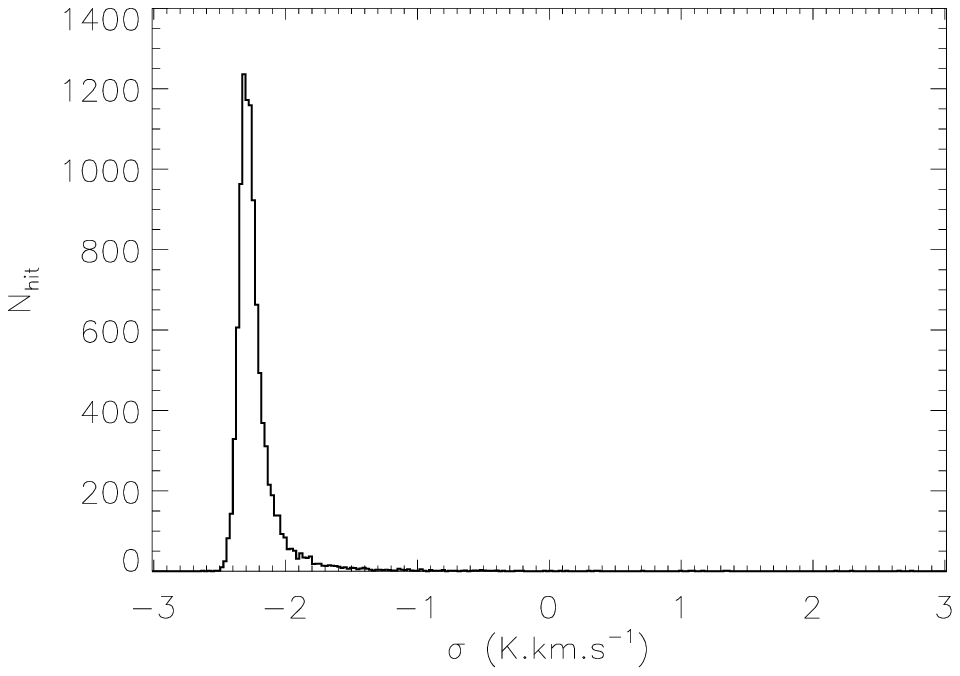}
\includegraphics[scale=0.25]{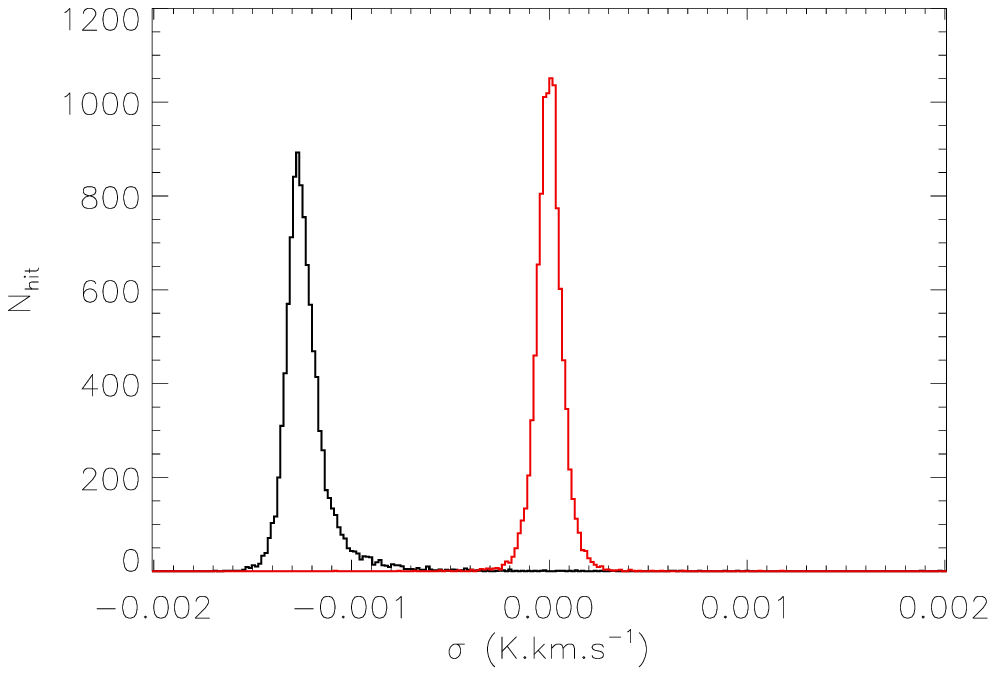}\\
\includegraphics[scale=0.25]{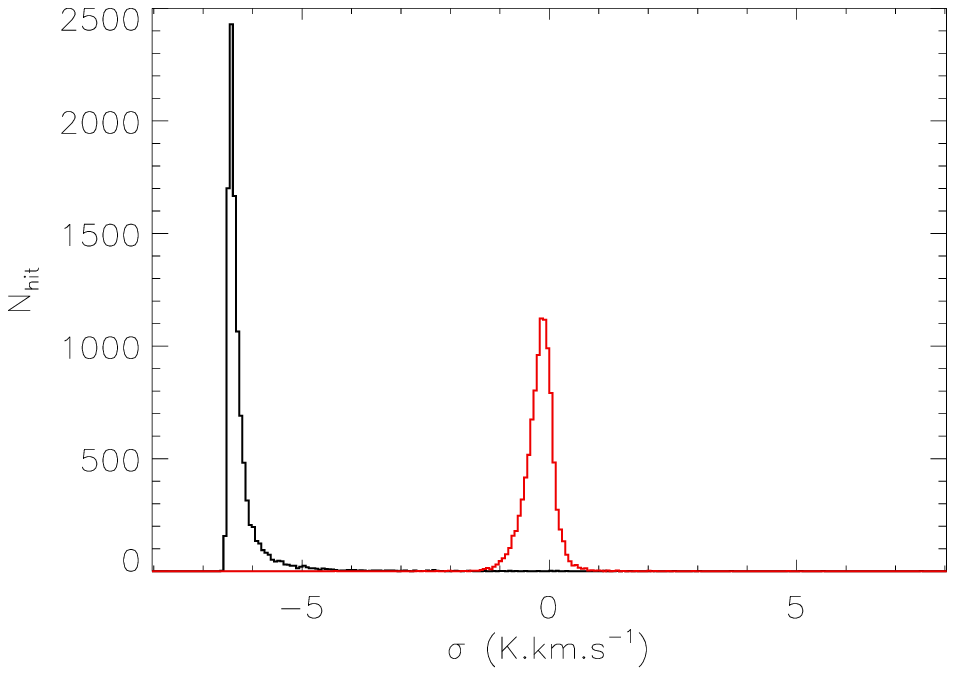}
\includegraphics[scale=0.25]{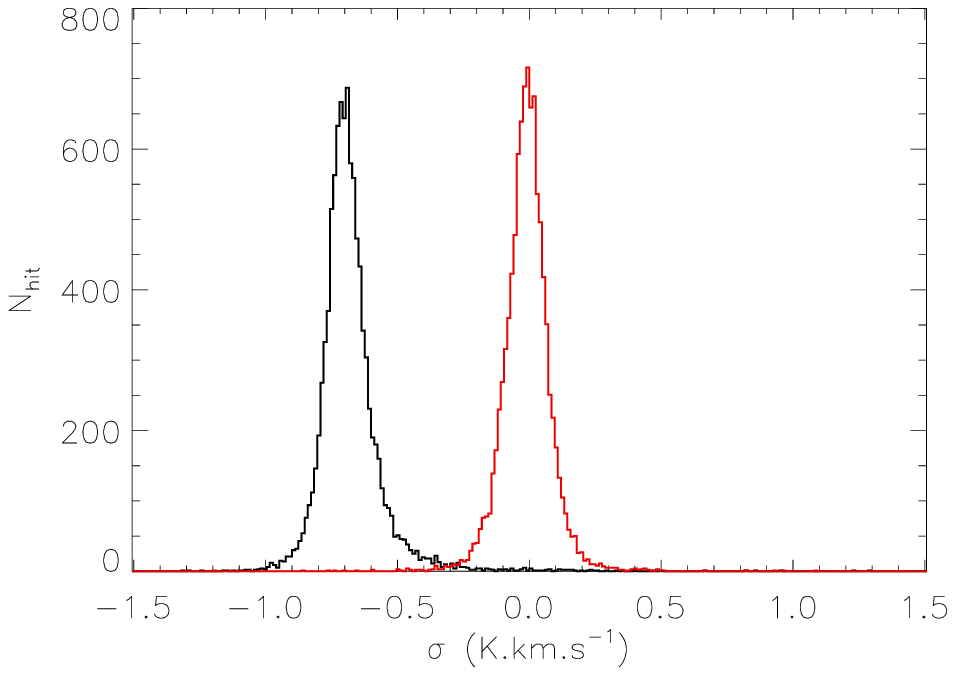}\\
\includegraphics[scale=0.25]{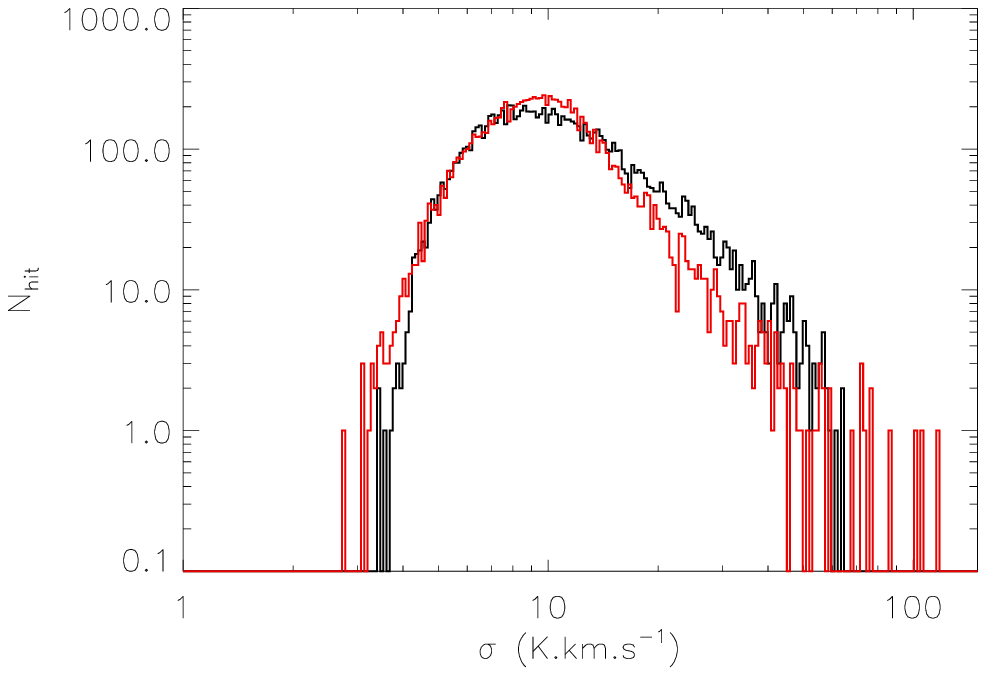}
\caption{From left to right and top to bottom : contamination by other components in CO reconstruction for 10 000 simulations for, CMB, tSZ, Dust + CIB, Synchrotron + Free-Free + Radio point sources and Noise respectively. In black for a standard CO adapted ILC and in red for MILCA. Each contamination are presented in terms of standard deviation in units of K.km.s$^{-1}$.}
\label{milcacosim}
\end{center}
\end{figure}

\section{CO emission reconstruction using the Planck bolometer maps}
\label{reconco}
We present  an original application of MILCA for reconstruction of  CO emission maps for the first three CO transitions
$(J=1-0)$, $(J=2-1)$ and $(J=3-2)$ using the spectral mismatch between the bolometers within a single Planck  channel
as discussed in \citet{planckco}.
We discuss here the methodology and performance of the algorithm using the simulations of the \textit{Planck} 100 GHz channel presented in Sect.~\ref{secsim}.
We consider four intensity maps constructed by combining the maps of two bolometers for each of the 4 PSB pairs.\\ 
Assuming bolometer maps are calibrated in CMB temperature units we can compute the bandpass mismatch for different astrophysical signals.
For CMB we expect no bandpass mismatch within the calibration uncertainties below 1 \%).
For other astrophysical components with a continuous emission spectrum, the variation of transmission in the different bolometers is very small (about 1\% at 100GHz).
However, for molecular line associated emission, such as CO, the transmission variations from one bolometer to another can reach up to 20\% \citep{planckco}. 
Notice that for CMB the transmission variations are null at the calibration uncertainties level. \\
We use this variation of transmission among bolometers to produce CO maps using MILCA. Formally speaking bandpass mismatch can be
accounted for by assuming a bolometer dependent mixing matrix (rather than one depending on observation frequency) and then component separation algorithms can
be applied. However, there are several problems with this approach. First, spectral bandpass mismatch is very sensitive to the
SED of the astrophysical component and it might be difficult to estimate for those astrophysical components for which the SED vary spatially.
Second, the signal to noise ratio is low with respect to conventional problems based on channel maps.
MILCA is well-adapted for this particular problem as it allow us on the one hand to only use the bandpass mismatch for those components for which is well known
(mainly CMB and the wanted component) and on the other hand to minimize the noise contribution in the final recovered map. \\

We performed 10 000 simulations of the \textit{Planck} bolometer maps  using different simulated bandpasses based on the CO unit conversion coefficients
values and uncertainties presented in \citet{planckco}.
Figure~\ref{milcaco} shows the comparison between the reconstructed CO emission at 100 GHz for an ILC algorithm spectrally adapted to CO (left) and for MILCA (right)
for one these simulations. The top panel shows the reconstructed CO map and the bottom panel the residuals in $K km s/s$ units.
We observe that the CO emission is well reconstructed with MILCA but in the inner Galactic plane where we observe some residual contamination.
However the ILC reconstructed map shows strong contamination by the CMB and the Galactic diffuse emission, and a larger noise level.
In Table \ref{tabco2} we summarize the contamination in terms of contribution to the variance of the CO reconstructed map for MILCA and for the CO adapted ILC. 
Notice that MILCA allows us to produce CO maps with a level of contamination 5 times lower than a CO adapted ILC.
At a resolution of 10 arcmin, the MILCA reconstructed CO map residual is dominated by the noise contribution, $6.06$~K.km/s over $6.07$~K.km/s, other components contribute only for $0.31$~K.km/s. For comparison in the ILC CO map the noise contribution is about $7.55$~K.km/s, and the contamination by other components reaches $5.53$~K.km/s.
Also notice that in the case of a MILCA reconstruction at 30 arcmin FWHM, the standard deviation of the map ($9.35$~K.km/s) is completely dominated by the CO emission (9.11~K.km/s) indicating the high level of purity of the map. For the ILC CO reconstructed map the standard deviation of the total map is $5.94$~K.km/s, which is below the CO only contribution ($9.11$~K.km/s. This illustrate the large bias in the ILC map for which the thermal dust emission is used to remove CO emission and then to reduce the total variance of the map.\\

Figure~\ref{milcacosim} presents the contamination level in the CO map for the 10 000 simulations for the ILC and the MILCA reconstructions. We represent on these figures the contamination by the CMB, the thermal dust + CIB, the tSZ  effect and the synchrotron + free-free + radio point sources and by the noise.\\
For CMB, the contamination in MILCA map is equal to zero at the calibration uncertainties level, but in the ILC map we observe a large contamination by the CMB, which accounts to 2.5~K.km/s in average in terms of the standard deviation. 
This contamination is mainly produced by the CO/galactic foregrounds correlation. Indeed the galactic foregrounds with a continuous emission law follow almost the same spectral dependence in the different bolometers than the CMB. Consequently, we observe similar residuals for all contaminating astrophysical emission at the percents level. The bias in the ILC map is mainly driven by the CO/thermal dust correlation, as the distribution of the dust contamination is the tightest.\\

\begin{table}
\center
\caption{Contribution to the total standard deviation of the reconstructed CO emission map in units of K.km/s}
\label{tabco2}
\begin{tabular}{|c|c|c|}
\hline
Maps & CO ILC ($10'$)& CO ILC ($30'$)\\
\hline
Total  & 10.43 & 5.94 \\
CO & 10.16 & 9.11 \\
Residuals & 9.36 & 4.64 \\
Noise & 7.55 & 0.57  \\
Other components & 5.53 & 4.60 \\
CMB & 1.45 & 1.14\\
Dust + CIB & 5.02 & 4.16  \\
Sync + Free-Free & 0.50 & 0.46  \\
SZ & 0.025 & 0.017 \\
\hline
Maps & CO MILCA ($10'$)& CO MILCA ($30'$)\\
\hline
Total  & 12.05 & 9.35\\
CO & 10.16 & 9.11\\
Residuals & 6.07 & 0.53 \\
Noise & 6.06 & 0.46 \\
Other components & 0.31 & 0.25 \\
CMB & 0 & 0 \\
Dust + CIB & 0.35 & 0.29 \\
Sync + Free-Free & 0.08 & 0.07 \\
SZ & 0.0020 & 0.0013 \\
\hline
\end{tabular}
\end{table}

\noindent MILCA have been applied to the \textit{Planck} bolometer maps at  100, 217 and 353 GHz in order to extract maps of the emission
of the first three CO lines J=1-0, J=2-1 and J=3-2 \citep{planckco}. These maps were released by the \textit{Planck} collaboration and 
have extensively been compared with external data in \citet{planckco}.

\section{Conclusion}

Component separation techniques are now a major ingredient in the scientific exploitation of 
multi-channel and multi-component datasets as those of the CMB satellite experiments \textit{WMAP} and \textit{Planck}. \\

In this paper we have presented the Modified Internal Linear Algorithm (MILCA) specially developed for the \textit{Planck} data.
MILCA generalizes the standard ILC algorithm, generally devoted to CMB emission extraction, to any astrophysical component with a known emission law. 
In practice MILCA can be used to extract an arbitrary number of astrophysical emissions rejecting the contribution from others for which the
emission law is also known. MILCA has been optimized in various ways in order to reduce significantly residuals from instrumental noise and
other astrophysical components. For this purpose the separation is performed both in real and Fourier spaces. Moreover
the data covariance matrix is modified and divided into multiple subspaces in order to avoid the confusion
between instrumental noise and astrophysical components in the original data. Finally, we have also introduced the possibility of using external
templates to improve the efficiency of the algorithm. Because of these improvements MILCA has been proved to be
more efficient than standard ILC algorithms in a large range of astrophysical problems. \\

We have applied MILCA to simulated \textit{Planck} data, and have shown that it can be used to efficiently reconstruct the tSZ and CO emissions.
We have proposed, with MILCA, an original way to used prior on the spatial distribution of contaminating components. Allowing to constraint emissions such as CIB, which can not be described with a single template and an SED.\\
In the tSZ case we have showed that MILCA can be used to reconstruct low signal to noise and highly non-Gaussian components in the \textit{Planck}
data. Indeed, MILCA has been used extensively by the \textit{Planck} collaboration for tSZ studies \citep{planck11d,planck11e,planck13b,planck13c,planck13e,planck2013-p05a,plancksz}. For CO emission we have presented a new component separation approach that takes advantage of the spectral bandpass mismatch between
bolometers of the same \textit{Planck} channel. We have proved that MILCA improvements on the reduction of the instrumental noise and unwanted astrophysical
components are critical for this. Maps of the emission for the first three CO transitions have been obtained by the \textit{Planck} collaboration using MILCA~\citep{planckco}.


\section*{Acknowledgements}
\thanks{We thank Nabila Aghanim and Fran\c{c}ois-Xavier D\'esert for useful 
discussions related to this work. 
Some of the results in this paper have been derived using the
\textit{HEALPix} package \citep{healpix}.}

\bibliographystyle{aa}
\bibliography{milca_v4}

\end{document}